\documentclass[%
twocolumn,
superscriptaddress,
amsmath,amssymb,
aps,
]{revtex4-2}
\usepackage{float}
\usepackage{graphicx}
\usepackage{dcolumn}
\usepackage{bm}
\usepackage[mathlines]{lineno}
\usepackage{tikz}
\usepackage{float}
\usepackage{subfigure}
\usepackage{verbatim}
\usepackage{epsfig}
\usepackage{epstopdf}
\usepackage{graphics}
\usepackage[utf8]{inputenc}
\usepackage{booktabs, array, mathptmx, float, tabularx, booktabs, lipsum, amsmath,multirow,braket}
\usepackage[colorlinks,linkcolor=blue,anchorcolor=blue,citecolor=blue]{hyperref}
\usepackage{mathtools,amsmath,graphicx,array,amssymb}
\usepackage[mathscr]{euscript}
\usepackage{dcolumn}
\usepackage{bm}
\usepackage{xcolor}
\usepackage{siunitx}
\usepackage{hyperref}
\hypersetup{colorlinks=true,
	linkcolor=blue,
	filecolor=blue,
	urlcolor=blue,
	citecolor=blue,
}
\usepackage{url}
\usepackage[utf8]{inputenc}
\usepackage[english]{babel}
\usepackage{upgreek}

\begin{document}


	\title{Variational Quantum Monte Carlo investigations of the superconducting pairing in La$_3$Ni$_2$O$_7$}


	\author{Yi-Qun Liu}
	\affiliation{National Laboratory of Solid State Microstructures $\&$ School of Physics, Nanjing University, Nanjing 210093, China}
	\author{Da Wang}
	\email{dawang@nju.edu.cn}
	\affiliation{National Laboratory of Solid State Microstructures $\&$ School of Physics, Nanjing University, Nanjing 210093, China}
	\affiliation{Collaborative Innovation Center of Advanced Microstructures, Nanjing University, Nanjing 210093, China}

	\author{Qiang-Hua Wang}
	\email{qhwang@nju.edu.cn}
	\affiliation{National Laboratory of Solid State Microstructures $\&$ School of Physics, Nanjing University, Nanjing 210093, China}
	\affiliation{Collaborative Innovation Center of Advanced Microstructures, Nanjing University, Nanjing 210093, China}



	\begin{abstract}
		We investigate the pairing symmetry in the novel superconductor La$_3$Ni$_2$O$_7$ under pressure by the non-perturbative variational quantum Monte Carlo. Within the bilayer Hubbard model and extended $t-J$ model with two orbitals in the $E_g$ doublet, we find the local strong correlation triggers $s_\pm$-wave Cooper pairing, with sign change of the gap function among the various Fermi pockets, while the $d_{x^2-y^2}$-wave pairing is generically disfavored. This is in agreement with the results from functional renormalization group applied in the weak up to moderate correlation limit. We find the 3d$_{3z^2-r^2}$ orbital plays a leading role in the superconducting pairing. We also demonstrate the finite intra-orbital double occupancy even in the strong correlation limit, shedding light on the itinerant versus local moment picture of the electrons in this material.
	\end{abstract}


	\maketitle


\section{Introduction}
	The discovery of high-temperature superconductivity in La$_3$Ni$_2$O$_7$ ($T_c\approx 80K$) under pressure \cite{Sun2023} triggers extensive research interests both experimentally \cite{zhang2023hightemperature,JunHou117302,liu2024electronic,wang2023pressureinduced,rhodes2023structural,yang2023orbitaldependent,zhou2023evidence,wang2023structure,kakoi2023multiband,wang2023longrange,Chen2024,dan2024spindensitywave,chen2024electronic,geisler2024optical,zhan2024cooperation} and theoretically \cite{PhysRevLett.131.126001,PhysRevB.108.125105,YangShen127401,PhysRevB.108.L201121,gu2023effective,sakakibara2024possible,PhysRevB.108.L140505,PhysRevB.108.L180510,PhysRevLett.132.036502,PhysRevB.108.174511,wu2023charge,PhysRevB.108.214522,zhang2023structural,lu2023interlayer,PhysRevLett.131.236002,chen2023critical,PhysRevB.109.L081105,PhysRevLett.131.206501,luo2023hightc,PhysRevLett.131.126001,jiang2023pressure,tian2023correlation,PhysRevB.108.L140504,PhysRevB.108.174501,Jiang_2024,PhysRevB.108.L201108,PhysRevB.109.045127,yang2023strong,ryee2023critical,lu2023interplay,PhysRevB.109.045154,gao2023la3ni2o65,ouyang2023hund,chen2023orbitalselective,chen2023evidence,schlomer2023superconductivity,zheng2023superconductivity,dong2023visualization,heier2023competing,kakoi2023pair,wang2024electronic,botzel2024theory,yi2024antiferromagnetic,ouyang2024absence,oh2024type}.
	The infinite-layer nickelate \cite{Li2019}, R$_3$Ni$_2$O$_7$
	(R is the rare-earth elements) \cite{pan2023effect,PhysRevB.108.165141},
	La$_4$Ni$_3$O$_{10}$ \cite{sakakibara2023theoretical,zhang2023superconductivity} and other candidates \cite{alvarez2023electronphonon} now stand as a new class of high-temperature superconductors besides copper oxide \cite{Bednorz1986}. In this paper, we concentrate on La$_3$Ni$_2$O$_7$. It has a bilayer structure, and the orbitals in two layers are hybridized by Ni-O-Ni bonding. This differs from cuprate materials, which are typically monolayer or multi-layer but the coupling between layers are weak  \cite{PhysRevB.50.438,PhysRevB.53.6786,PhysRevB.52.7708}. In addition to this, both the 3d$_{3z^2-r^2}$ (nearly half-filled) and 3d$_{x^2-y^2}$ (nearly quarter-filled) orbitals in the E$_g$-multiplets in La$_3$Ni$_2$O$_7$ are active near the Fermi level under high pressure \cite{Sun2023,yang2023orbitaldependent,PhysRevLett.131.126001,PhysRevB.108.L180510} while the cuprate can be treated as a single-orbital system.
	There are two important questions to ask for the microscopic understanding of superconductivity in La$_3$Ni$_2$O$_7$. The first issue is the pairing symmetry. The majority of theories suggest $s_{\pm}$-wave pairing, with sign change of the gap function among the various Fermi pockets \cite{PhysRevB.108.L140505,gu2023effective,sakakibara2024possible,PhysRevLett.132.036502,PhysRevB.108.174511,zhang2023structural,lu2023interlayer,PhysRevLett.131.236002,luo2023hightc,tian2023correlation,PhysRevB.108.L140504,PhysRevB.108.174501}, while there are also theories suggesting $d_{x^2-y^2}$-wave pairing \cite{jiang2023pressure,Jiang_2024,wang2024electronic,heier2023competing} or competition between the two cases of pairing symmetry \cite{PhysRevB.108.214522}. The other issue is which orbital plays the dominant role for superconductivity? The viewpoint differs: Either the 3d$_{3z^2-r^2}$ orbital dominates \cite{PhysRevLett.131.206501,gu2023effective,PhysRevB.108.L140505,wu2023charge,zhang2023structural,luo2023hightc,PhysRevB.108.174501,PhysRevB.109.045154}, or the 3d$_{x^2-y^2}$ orbital plays the primary role \cite{PhysRevLett.132.036502,PhysRevB.108.174511,tian2023correlation,lu2023interplay,chen2023orbitalselective}. 

	La$_3$Ni$_2$O$_7$ has been extensively studied within the bilayer two-orbital model by various approaches, such as functional renormalization group (FRG) and renormalized mean field theory (RMFT) \cite{PhysRevB.108.L140505}, random phase approximation (RPA) \cite{PhysRevLett.131.236002}, dynamical mean field theory (DMFT) \cite{wang2024electronic}, density matrix renormalization group (DMRG) \cite{kakoi2023pair,schlomer2023superconductivity}, infinite projected entangled pair states (iPEPS) \cite{chen2023orbitalselective} and slave boson mean field (SBMF) \cite{lu2023interplay,Jiang_2024}, etc. Each approach has its own advantage and limitations. For example, the FRG and RPA work in the itinerant picture, hopefully applicable in the weak up to moderate correlation regime. The various mean field theories are non-perturbative but suffer from the known problem of ignoring the correlation of fluctuations. The DMFT takes local quantum fluctuations into account but yet the spatial correlations are still unaccessible. The DMRG and iPEPS studies are theoretically exact, but are limited to finite size or finite bond dimension. Moreover, it is not yet clear whether the starting model could be defined in the local moment limit, and even in this limit whether the local moment on the $d_{z^2}$ orbital could be effectively dropt out, assuming the Hund's coupling is also strong. In this work, we use the variational quantum Monte Carlo (VQMC) simulations to study the bilayer Hubbard model and its low energy effective extended $t-J$ model (allowing orbital double occupancy). Within the variational ansatz, the VQMC is able to handle a relatively larger lattice size, and to treat the correlation effects exactly. We compare the optimized ground state energies in the $s_{\pm}$-wave pairing ansatz and the $d_{x^2-y^2}$-wave pairing one. We find that, for the Hubbard model, the $s_{\pm}$-wave pairing symmetry is generically more favorable for not too weak Hund's coupling, and for the extended $t-J$ model, the $s_{\pm}$-wave pairing always wins. Inspection of the variational pairing amplitudes ascertains that the 3d$_{3z^2-r^2}$ orbital predominates in the superconducting pairing. The results are consistent with the conclusion from the FRG study \cite{PhysRevB.108.L140505}. Moreover, we demonstrate sizable intra-orbital double occupancy in both models up to strong correlation limit, seeding light on the itinerant versus the local moment picture for the electrons in the superconducting La$_3$Ni$_2$O$_7$.

	The rest of the paper is arranged as follows: We first introduce the two-orbital Hubbard model and its effective $t-J$ model in strong coupling limit in Sec. II. Then, we employ VQMC to study and compare these two models in Sec. III. Finally, a summary is given in Sec. IV.


\section{Model}
\begin{figure}
\begin{tikzpicture}
			\node[anchor=south west,inner sep=0] (image) at (0,0) {\subfigure{\includegraphics[trim=10 0 10 10 ,clip,width=0.21\textwidth]{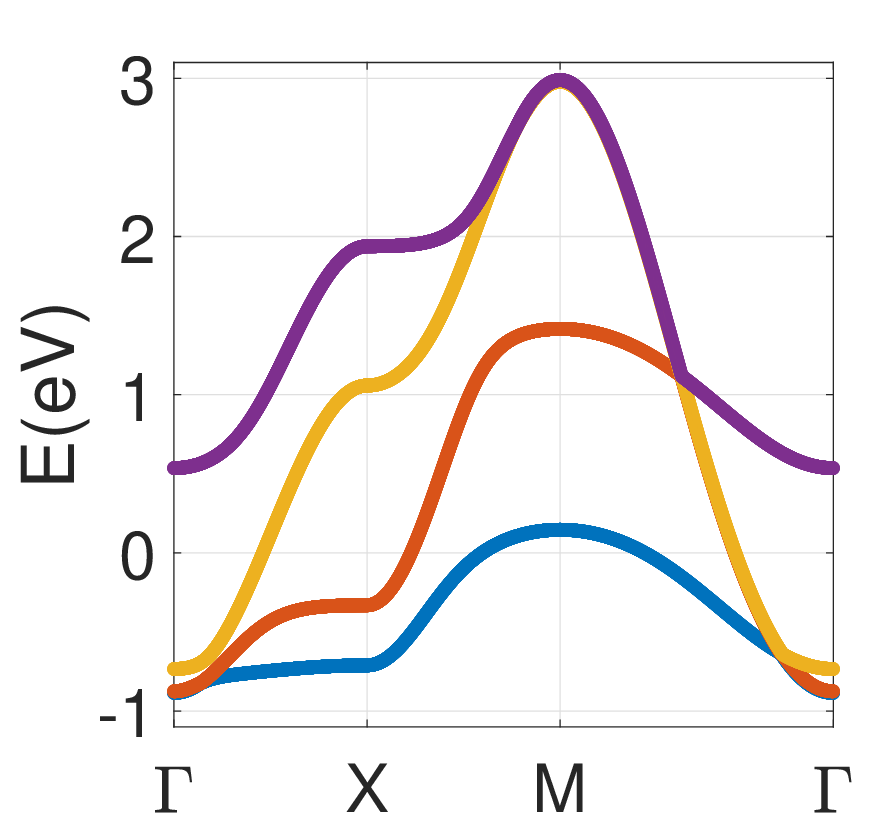}\label{fig:band}}};
			\node[above left] at (0.25,3.2) {(a)};
		\end{tikzpicture}
		\begin{tikzpicture}

			\node[anchor=south west,inner sep=0] (image) at (0,0) {\subfigure{\includegraphics[trim=0 30 0 10,clip,width=0.23\textwidth]{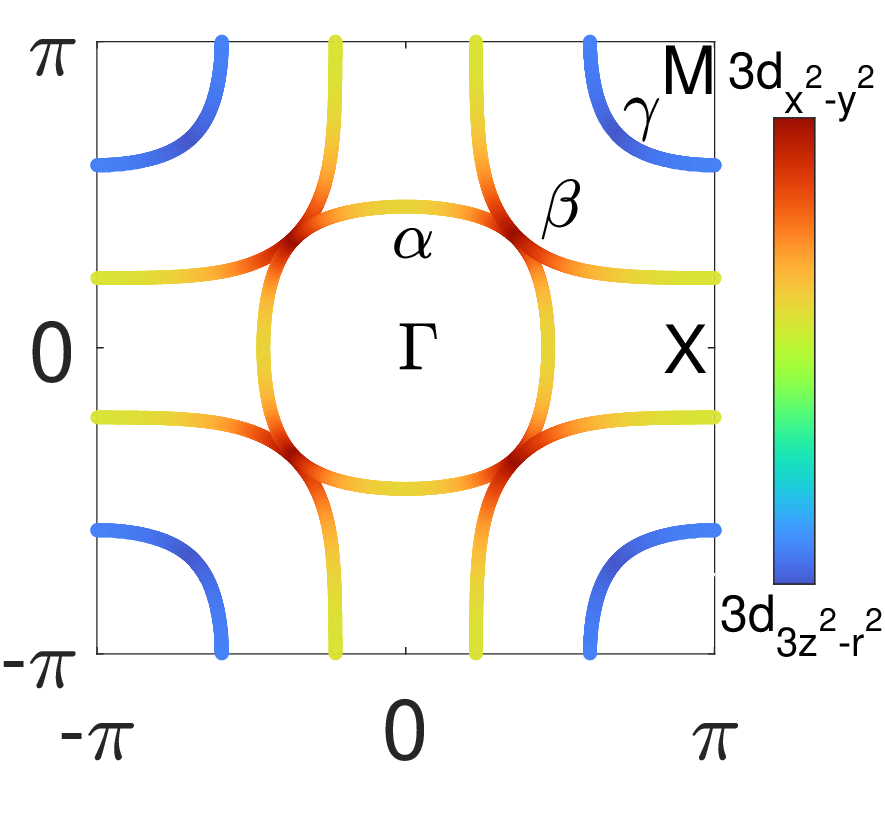}\label{fig:fermi}}};
			\node[above left] at (0.25,3.2) {(b)};

		\end{tikzpicture}

		\caption{ (a) Energy band dispersions along the high-symmetry path in the Brillouin zone. (b) Fermi surfaces with orbital components denoted by colors. The three pockets are labeled by $\alpha$, $\beta$ and $\gamma$, respectively. }
		\label{fig:1}
\end{figure}

To describe the band structure of La$_3$Ni$_2$O$_7$, we adopt the two-orbital (3d$_{x^2-y^2}$ and 3d$_{3z^2-r^2}$) tight-binding model on a bilayer square lattice given by the following Hamiltonian
\begin{align}
H_0=\sum_{i\delta,ab,\sigma}t_{\delta}^{ab}c_{ia\sigma}^{\dagger}c_{i+\delta b\sigma}+\sum_{ia\sigma}\varepsilon_ac_{ia\sigma}^{\dagger}c_{ia\sigma} ,
\end{align}
where $c_{ia\sigma}^\dag$ creates an electron on site $i$ with the orbital $a=x$ (for 3d$_{x^2-y^2}$) or $z$ (for 3d$_{3z^2-r^2}$) and spin $\sigma$. The matrix element $t_\delta^{ab}$ is the hopping integral between $a$ orbital on site $i$ and $b$ orbital on site $i+\delta$, and $\varepsilon_a$ denotes the onsite energy of the $a$ orbital. The tight-binding parameters are taken from Ref.~\cite{PhysRevLett.131.126001}: ($t_{100}^{xx}$, $t_{110}^{xx}$, $t_{00\frac{1}{2}}^{xx}$, $t_{100}^{zz}$, $t_{110}^{zz}$, $t_{00\frac{1}{2}}^{zz}$, $t_{100}^{xz}$, $t_{10\frac{1}{2}}^{xz}$, $\varepsilon_x$, $\varepsilon_z$) = ($-0.483$, $0.069$, $0.005$, $-0.110$, $-0.017$, $-0.635$, $0.239$, $-0.034$, $0.776$, $0.409$)~eV up to $C_{4v}$ operations. Here, the distance between layers is assigned as $\frac12$. The energy band dispersions and Fermi surfaces are shown in Fig.~\ref{fig:1} with the orbital characters denoted by colors (red for 3d$_{x^2-y^2}$ and blue for 3d$_{3z^2-r^2}$). There are three Fermi pockets: $\alpha$ (electron-like pocket around $\Gamma$), $\beta$ and $\gamma$ (hole-like pockets around $M$). Note that the $\gamma$-pocket around $M$ is mainly comprised of the 3d$_{3z^2-r^2}$ orbital, suggesting considerable pairing component on the 3d$_{3z^2-r^2}$ orbital, otherwise no pairing energy can be saved on the $\gamma$-pocket.

\begin{figure}
\begin{tikzpicture}
			\node[anchor=south west,inner sep=0] (image) at (0,0) {\subfigure{\includegraphics[trim=60 60 60 30 ,clip,width=0.22\textwidth]{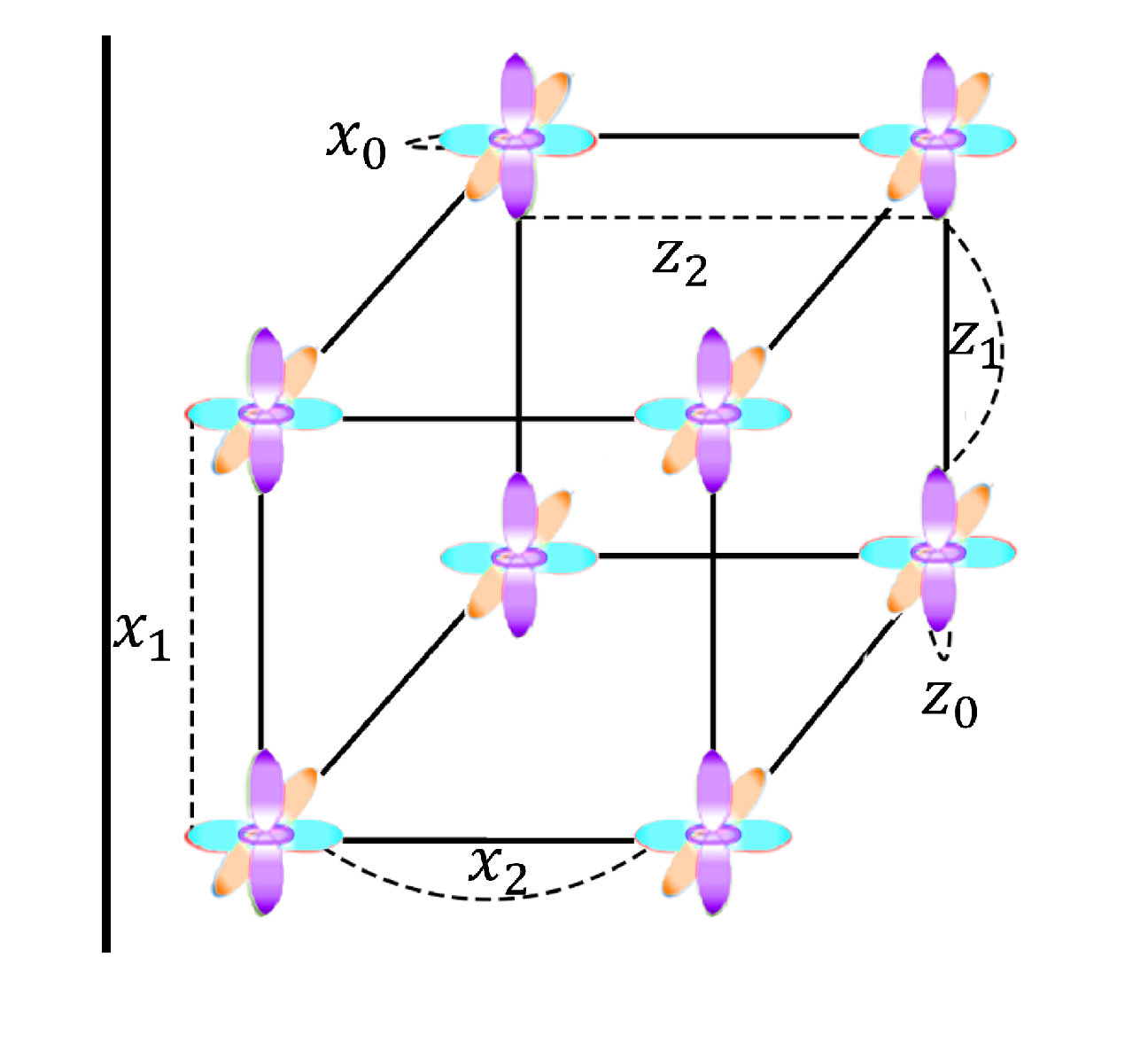}\label{fig:conswave}}};
			\node[above left] at (0.5,3.2) {(a)};
		\end{tikzpicture}
		\begin{tikzpicture}

			\node[anchor=south west,inner sep=0] (image) at (0,0) {\subfigure{\includegraphics[trim=60 60 60 30,clip,width=0.22\textwidth]{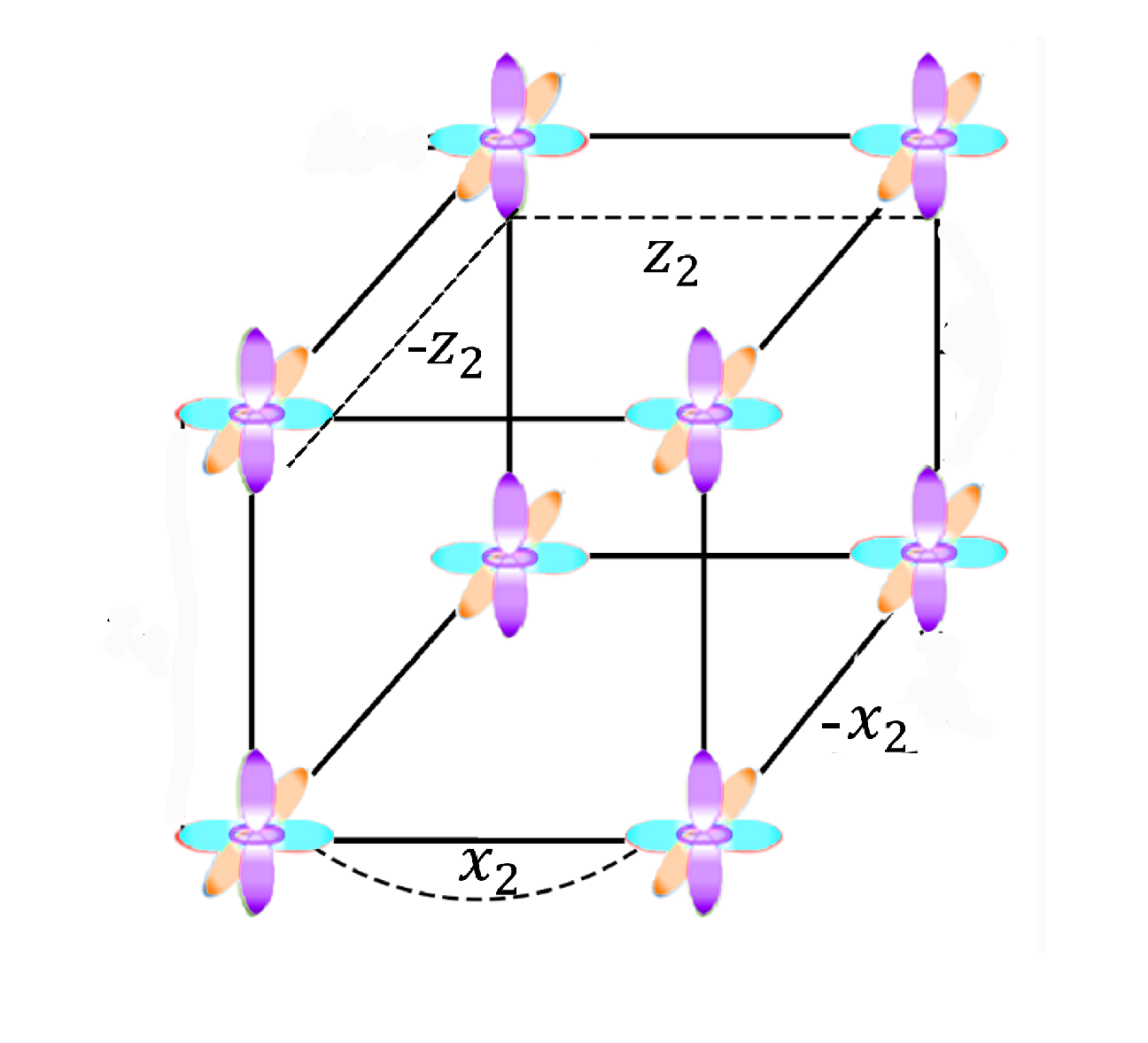}\label{fig:condwave}}};
			\node[above left] at (0.5,3.2) {(b)};

		\end{tikzpicture}
		\begin{tikzpicture}
			\vspace{-10pt}
			\node[anchor=south west,inner sep=0] (image) at (0,0) {\subfigure{\includegraphics[trim=10 10 10 10 ,clip,width=0.21\textwidth]{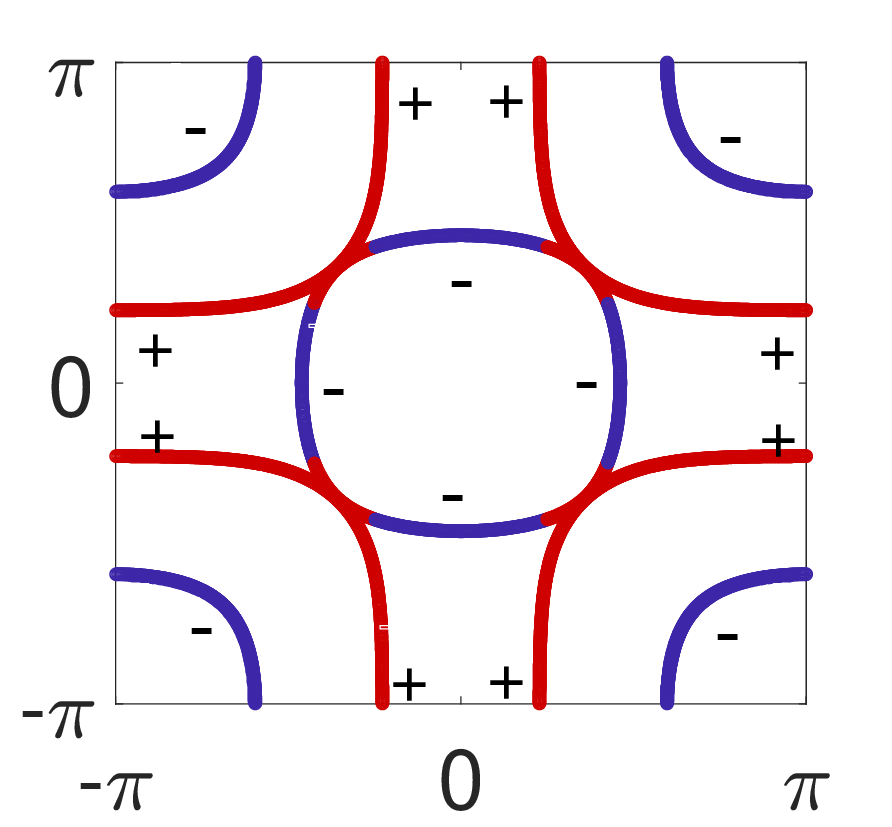}\label{fig:swave}}};
			\node[above left] at (0.25,3.2) {(c)};
		\end{tikzpicture}
		\begin{tikzpicture}

			\node[anchor=south west,inner sep=0] (image) at (0,0) {\subfigure{\includegraphics[trim=10 10 10 10,clip,width=0.21\textwidth]{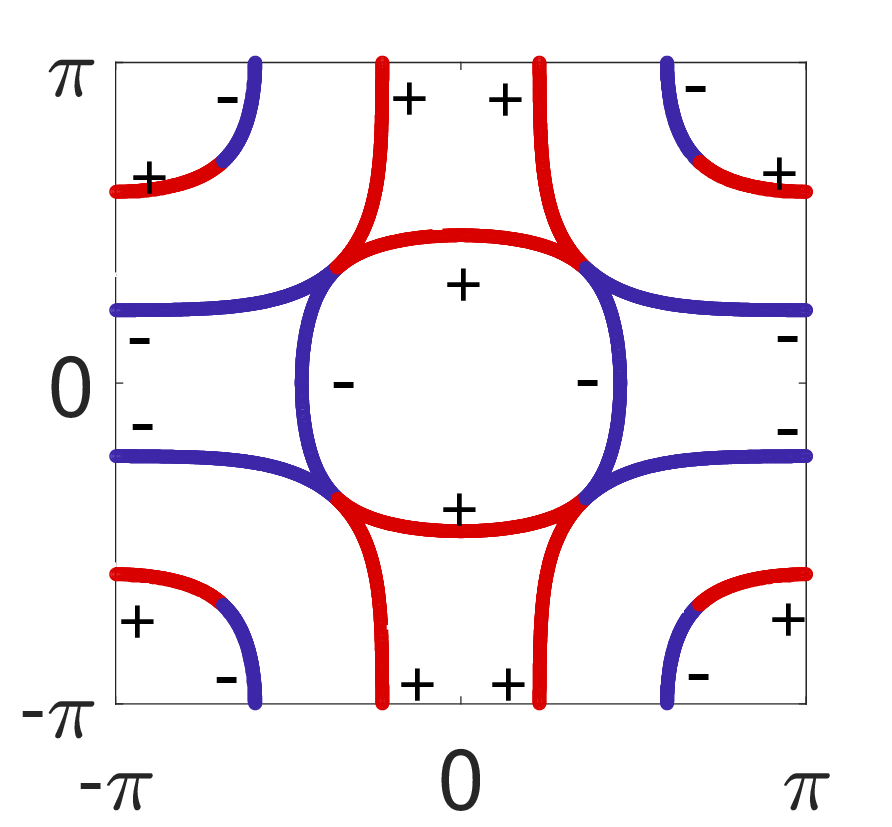}\label{fig:dwave}}};
			\node[above left] at (0.25,3.2) {(d)};

		\end{tikzpicture}
		\caption{ (a) Schematic plot of the pairing potentials (or variational pairing parameters in VQMC) for the $s_{\pm}$-wave pairing. Two orbitals, 3$d_{3z^2-r^2}$ and 3$d_{x^2-y^2}$, are assigned on each site. The dominant pairings between them are represented by $z_{0,1,2}$ and $x_{0,1,2}$, respectively. Here, $0$ denotes onsite paring, $1$ denotes the vertical bond, $2$ denotes the planar nearest neighbor bond. (b) is similar to (a) but for $d_{x^2-y^2}$-wave pairing. In (c) and (d), the typical gap functions for the $s_{\pm}$-wave and $d_{x^2-y^2}$-wave pairings are plotted on the Fermi surfaces with the sign indicated by colors (red for positive and blue for negative).}
		\label{fig:2}
\end{figure}
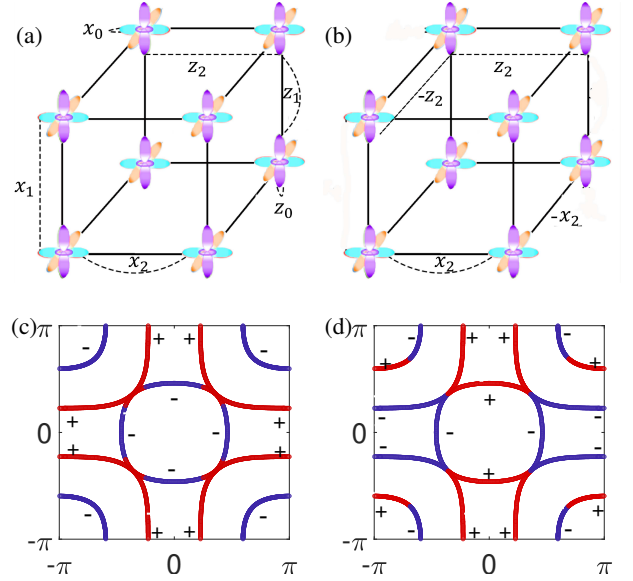

We consider the correlation effect by adding the onsite two-orbital Hubbard interactions given by
\begin{align} \label{Hubbard}
	H_H &=H_0+ \frac{1}{2}\sum_{i,a\neq b,\sigma\sigma^{\prime}}(U^{\prime}n_{ia\sigma}n_{ib\sigma^{\prime}}+J_Hc_{ia\sigma}^{\dagger}c_{ib\sigma}c_{ib\sigma^{\prime}}^{\dagger}c_{ia\sigma^{\prime}})\nonumber\\
	&+\sum_{ia}Un_{ia\uparrow}n_{ia\downarrow}+\sum_{i,a\neq b}J_Pc_{ia\uparrow}^{\dagger}c_{ia\downarrow}^{\dagger}c_{ib\downarrow}c_{ib\uparrow},
\end{align}
where $U$/$U'$ are intra/inter-orbital Coulomb repulsion, $J_H$ is the Hund's coupling and $J_P$ denotes the pair-hopping process. These four terms are assumed to satisfy the Kanamori relations \cite{PhysRevB.18.4945}: $U=U'+2J_H$ and $J_H=J_P$. Clearly, the bare Coulomb interactions are repulsive in the pairing channel and one needs to go beyond the bare interaction level to obtain superconductivity.
For weak couplings, the spin/charge/pair fluctuations are captured by FRG on equal footing, which predicts the s$_\pm$-wave pairing \cite{PhysRevB.108.L140505}. The dominant pairing terms are shown in Fig.~\ref{fig:conswave}, including onsite pairing denoted by $z_0$/$x_0$, pairing on the vertical bond $z_1$/$x_1$, and pairing on the in-plane nearest neighbor bond $z_2$/$x_2$. A typical gap function projected on the Fermi surfaces is shown in Fig.~\ref{fig:swave} with different signs on different pockets, hence, called s$_\pm$-wave pairing. As a comparison, for d$_{x^2-y^2}$-wave pairing as the subleading instability in FRG, the real space pairing terms denoted by $x_2$/$z_2$ are sketched in Fig.~\ref{fig:condwave} and the gap function on the Fermi surfaces is plotted in Fig.~\ref{fig:dwave}.

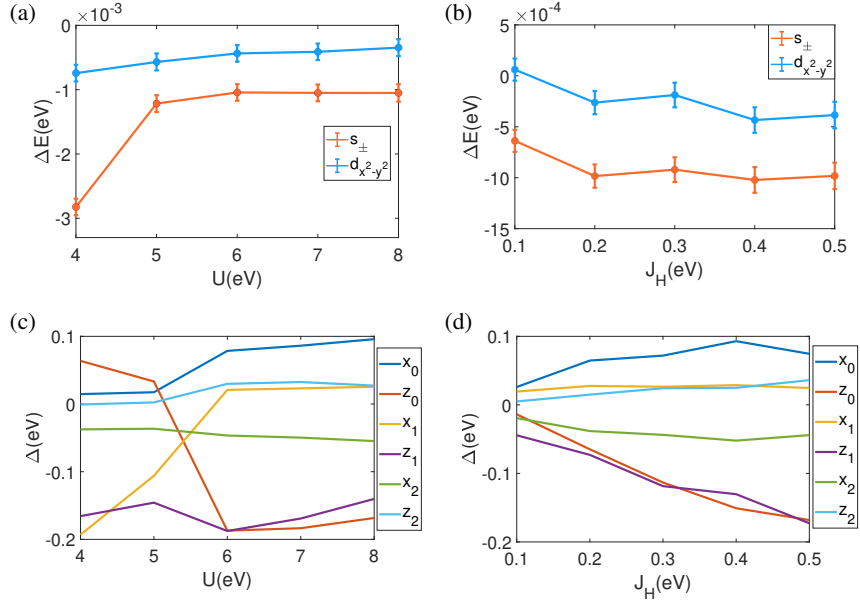
\begin{figure*}
		\begin{tikzpicture}
		\node[anchor=south west,inner sep=0] (image) at (0,0) {\subfigure{\includegraphics[trim=6 0 10 5 ,clip,width=0.3\textwidth]{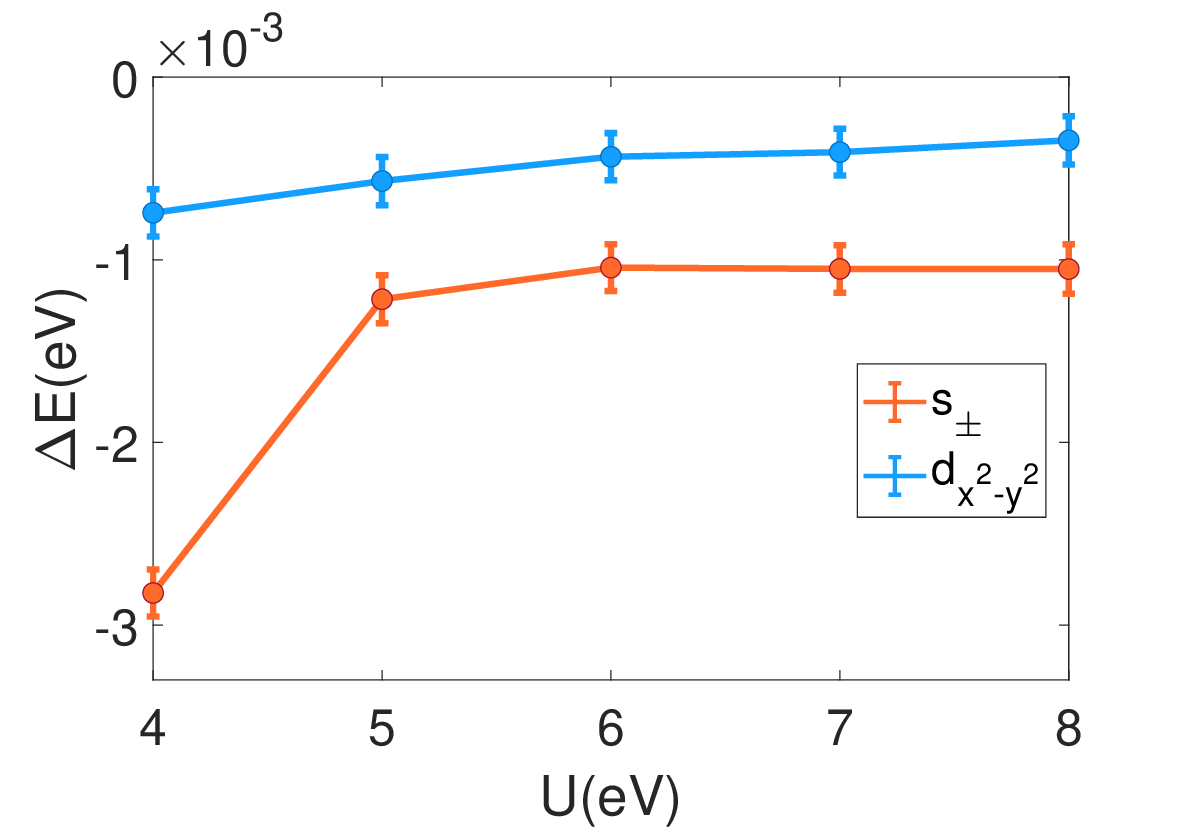}\label{fig:energyU}}};
		\node[above left] at (0.25,3.6) {(a)};
	\end{tikzpicture}
	\begin{tikzpicture}

		\node[anchor=south west,inner sep=0] (image) at (0,0) {\subfigure{\includegraphics[trim=7 0 10 5,clip,width=0.3\textwidth]{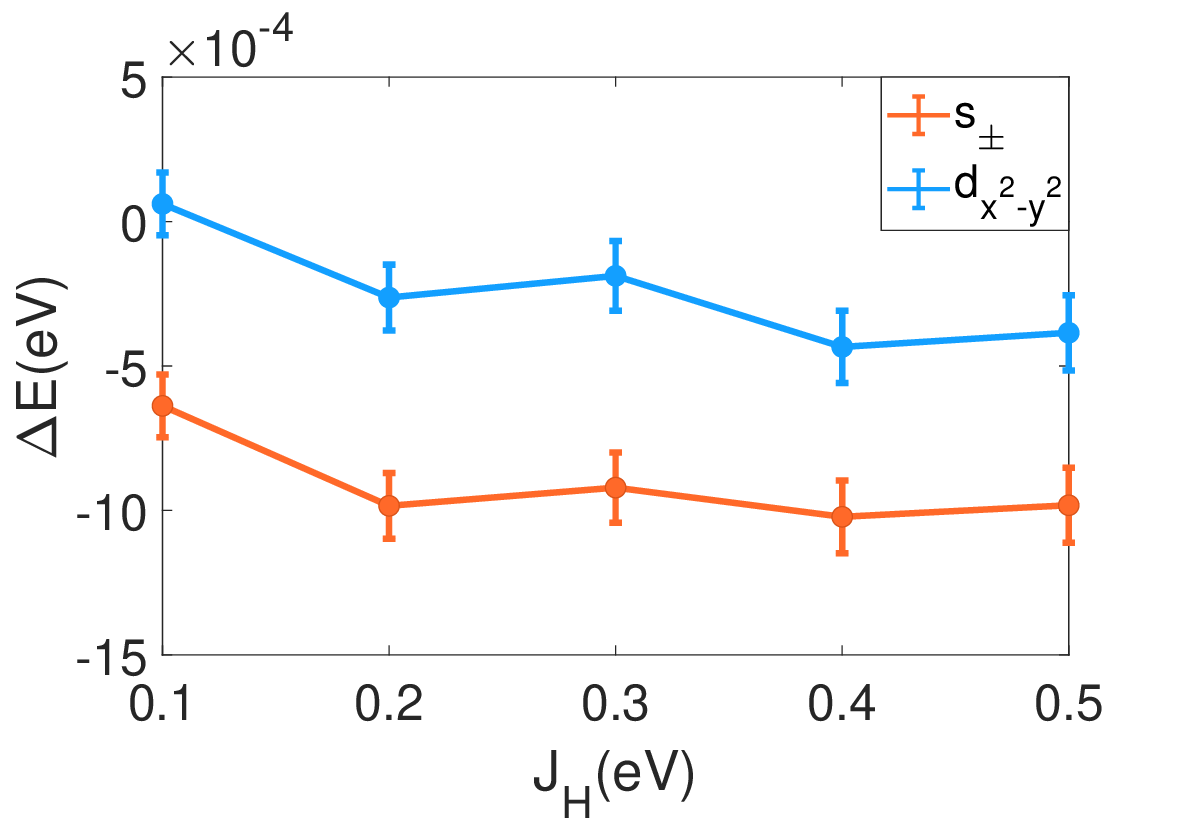}\label{fig:energyjh}}};
		\node[above left] at (0.25,3.6) {(b)};
	\end{tikzpicture}
	\\
	\begin{tikzpicture}
		\node[anchor=south west,inner sep=0] (image) at (0,0) {\subfigure{\includegraphics[trim=10 5 0 10 ,clip,width=0.3\textwidth]{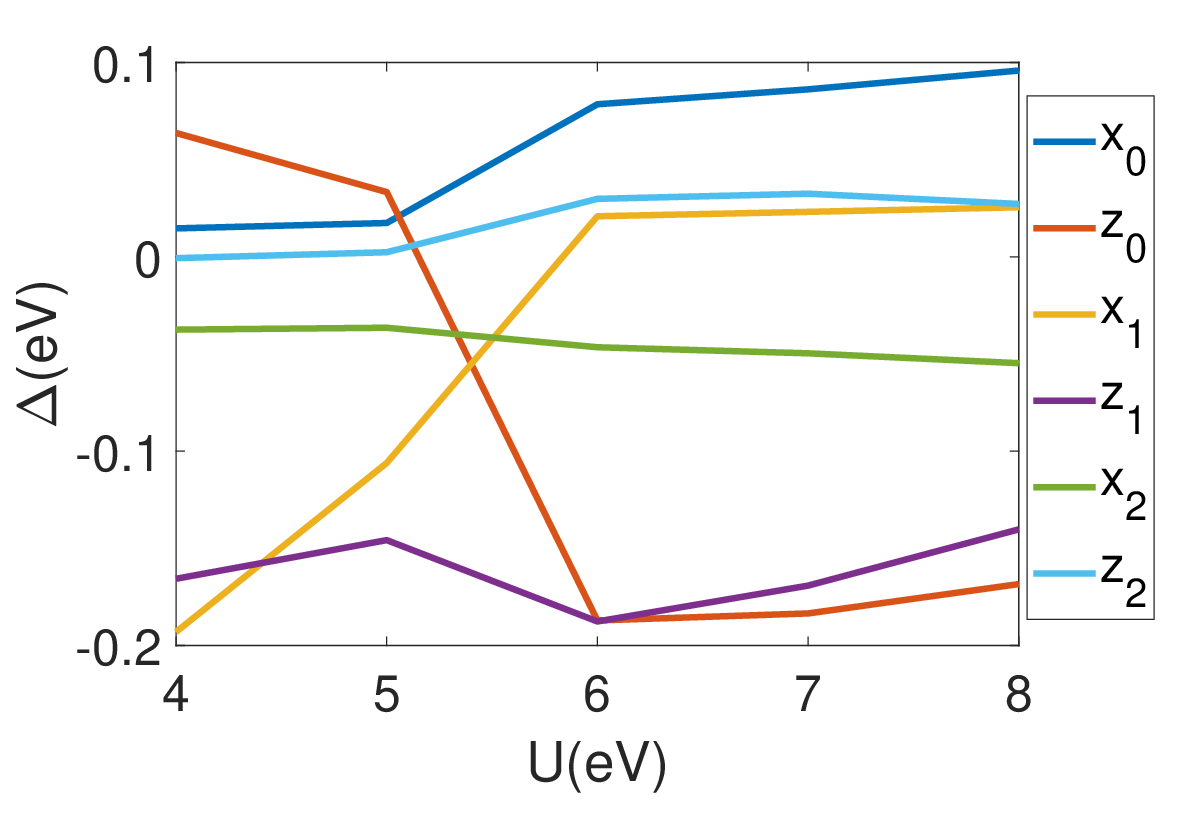}\label{fig:orderU}}};
		\node[above left] at (0.25,3.6) {(c)};
	\end{tikzpicture}
	\begin{tikzpicture}

		\node[anchor=south west,inner sep=0] (image) at (0,0) {\subfigure{\includegraphics[trim=10 5 0 10,clip,width=0.3\textwidth]{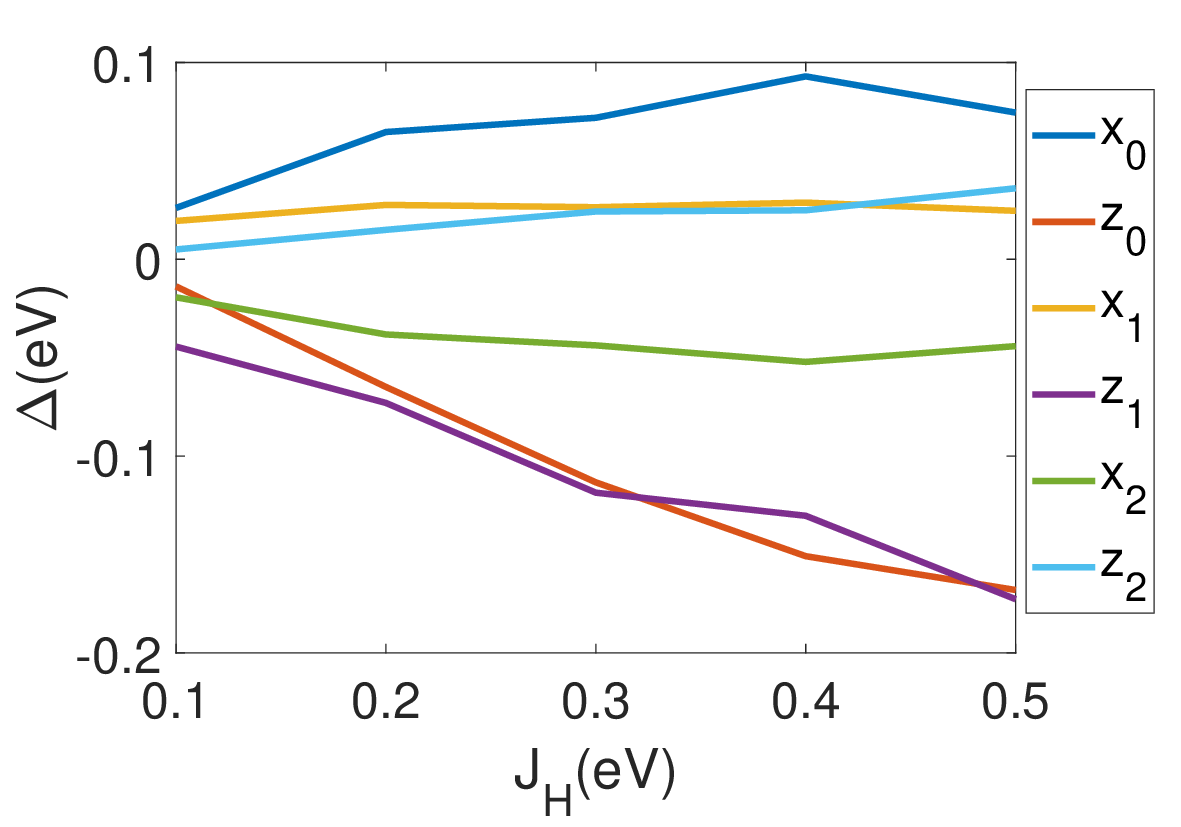}\label{fig:orderjh}}};
		\node[above left] at (0.25,3.6) {(d)};

	\end{tikzpicture}
\caption{VQMC results of the $t-J$ model. The energies of the $s_\pm$-wave and $d_{x^2-y^2}$-wave pairings relative to the normal state are plotted versus $U$ by fixing $J_H=0.5$~eV in (a), and versus $J_H$ by fixing $U=7$~eV in (b). The corresponding optimized variational pairing parameters for the $s_\pm$-wave are plotted versus $U$ in (c) and versus $J_H$ in (d), respectively.
\label{fig:tj}}
\end{figure*}

On the other hand, in the strong coupling limit with $U\gtrsim U'\gg J_H$, the above Hubbard model can be mapped to a two-orbital $t-J$ model \cite{PhysRevB.108.L140505} in the subspace of one-electron (1e) and two-electrons (2e) per site since there are $1.5$ electrons per site on average for La$_3$Ni$_2$O$_7$. The resulting $t-J$ Hamiltonian is
\begin{align} \label{t-J}
	H_{tJ} =& \sum_{\langle i j\rangle a, b, \sigma}\left(t_{i j}^{a b} c_{i a \sigma}^{\dagger} c_{j b \sigma} Q_{1 e, i} Q_{2 e, j}+\text { h.c. }\right) \nonumber\\
	& +\sum_{\langle i j\rangle a, n=1,2} J_{i j}^{a}\left(\mathbf{S}_{i a} \cdot \mathbf{S}_{j a}-\frac{1}{4} n_{i a} n_{j a}\right) Q_{n e, i} Q_{n e, j} \nonumber\\
	& +\sum_{i, a} J_{H}\left(Q_{2 e, i a}-Q_{1 e, i a} Q_{1 e, i \bar{a}}\right) \nonumber\\
	& +\sum_{i, a>b} J_{H}\left(\frac{1}{4} n_{i a} n_{i b}-\mathbf{S}_{i a} \cdot \mathbf{S}_{i b}\right) Q_{1 e, i a} Q_{1 e, i b} \nonumber\\
	& +\sum_{i, a>b} J_{P}\left(c_{i a \uparrow}^{\dagger} c_{i a \downarrow}^{\dagger} c_{i b \downarrow} c_{i b \uparrow}+\text { h.c. }\right) \nonumber\\
	& +\sum_{i a} \frac12(\varepsilon_{a}-\varepsilon_{\bar{a}})\left(Q_{1 e, i a}+2 Q_{2 e, i a}\right) ,
\end{align}
where $\bar{a}$ denotes the orbital other than $a$, $\mathbf{S}_{ia}$ and $n_{ia}$ are standard spin and charge operators for orbital $a$ on site $i$,
$Q_{ne,ia}$ is the projection operator for $n$-electrons on orbital $a$ and site $i$,
\begin{align}
Q_{0e,ia}&=(1-n_{ia\uparrow})(1-n_{ia\downarrow}), \quad
Q_{2e,ia}=n_{ia\uparrow}n_{ia\downarrow}, \nonumber\\
Q_{1e,ia}&=n_{ia\uparrow}(1-n_{ia\downarrow})+(1-n_{ia\uparrow})n_{ia\downarrow},
\end{align}
and $Q_{ne,i}$ is the projection operator for total charge $n$-electrons on site $i$,
\begin{align}
Q_{ne,i}=\sum_{k=0}^nQ_{ke,ix}Q_{(n-k)e,iz} .
\end{align}
The second line in Eq.~\ref{t-J} is the super-exchange (with $3e$ or $0e$ states as the intermediate states) $J_{ij}^a=4(t_{ij}^{aa})^2/\overline{U}$ with $\overline{U}=(U+U^{\prime})/2=U-J_H$. The inter-orbital super-exchange is neglected due to small values of $t_{ij}^{a\bar{a}}$ and $t_{ij}^{aa}t_{ji}^{\bar{a}\bar{a}}$.
Note that the above intra-orbital super-exchange term favors spin-singlet pairing at the mean field level, reminiscent of the single-orbital $t-J$ model for cuprates.
Since the largest super-exchange is between the 3d$_{3z^2-r^2}$ orbitals on the vertical bond, the $s_\pm$-pairing is expected to be favored with the dominant pairing component between the 3d$_{3z^2-r^2}$ orbitals on this bond, as indicated by $z_1$ in Fig.~\ref{fig:conswave}.
This is actually confirmed by previous RMFT calculations \cite{PhysRevB.108.L140505}.
In the following, we further employ VQMC to investigate both the above $t-J$ and Hubbard models.


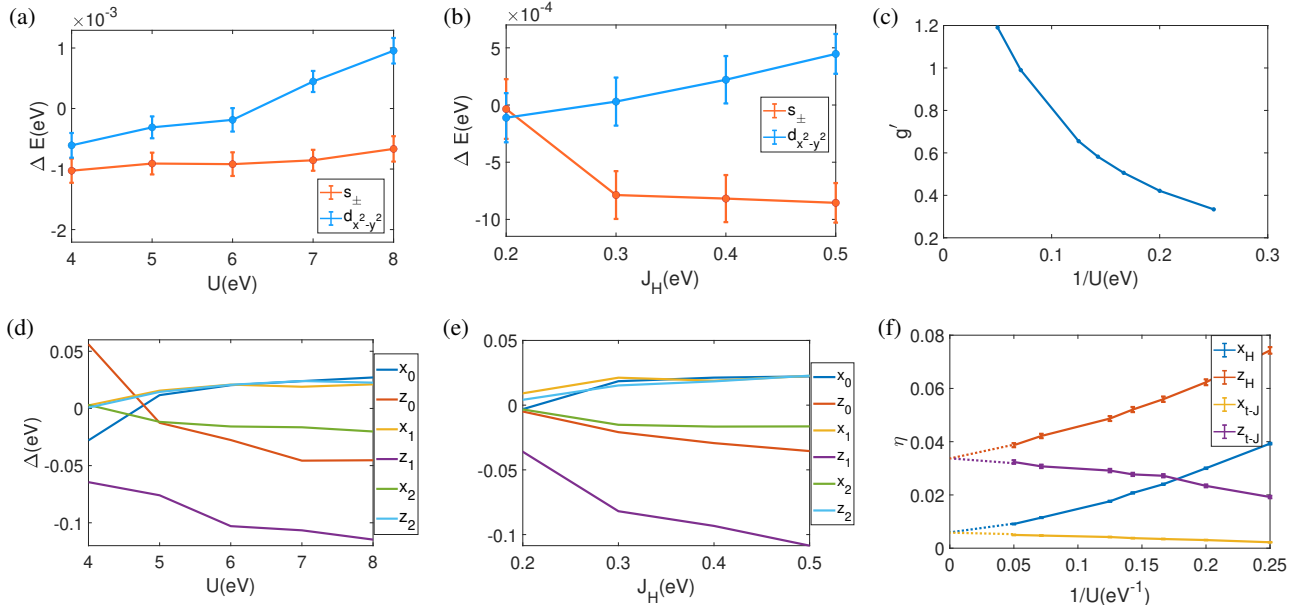
\begin{figure*}
		\begin{tikzpicture}
	\node[anchor=south west,inner sep=0] (image) at (0,0) {\subfigure{\includegraphics[trim=10 0 5 5 ,clip,width=0.3\textwidth]{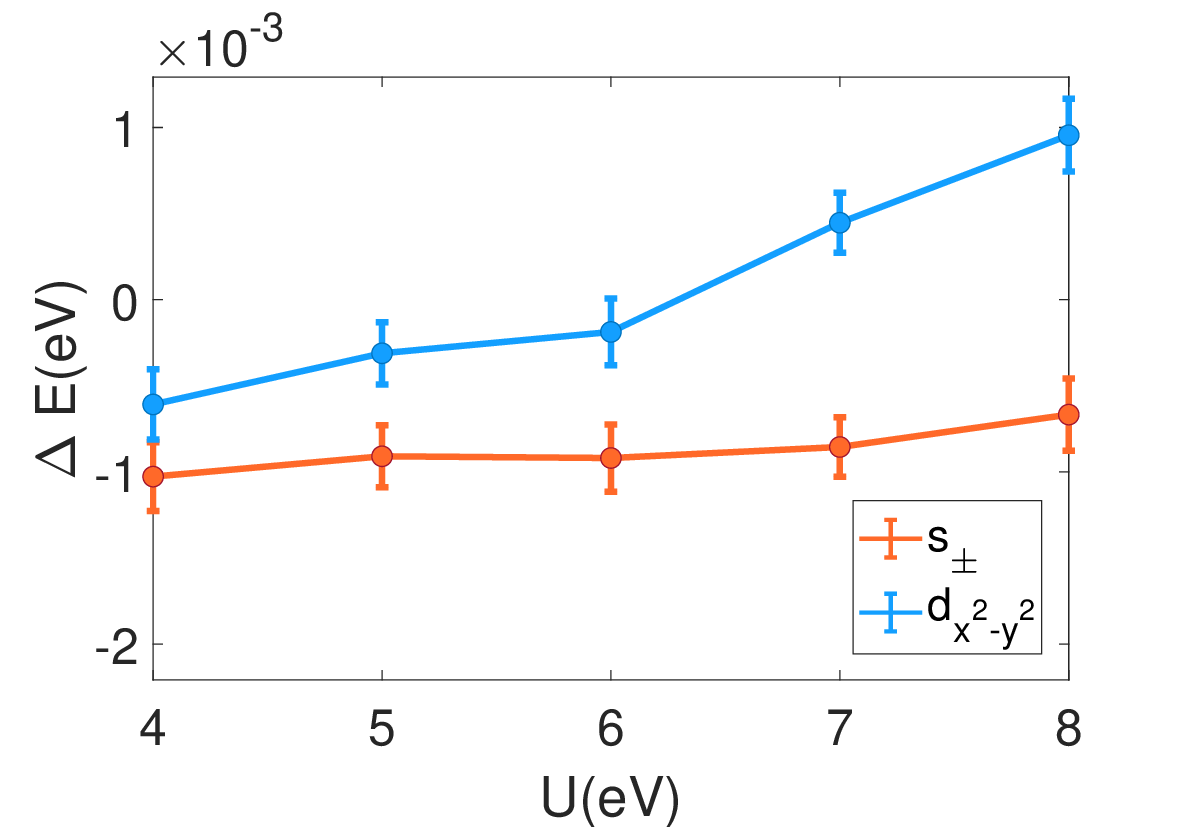}\label{fig:energyUhubbard}}};
	\node[above left] at (0.26,3.6) {(a)};
\end{tikzpicture}
\begin{tikzpicture}
	\node[anchor=south west,inner sep=0] (image) at (0,0) {\subfigure{\includegraphics[trim=8 5 24 5,clip,width=0.3\textwidth]{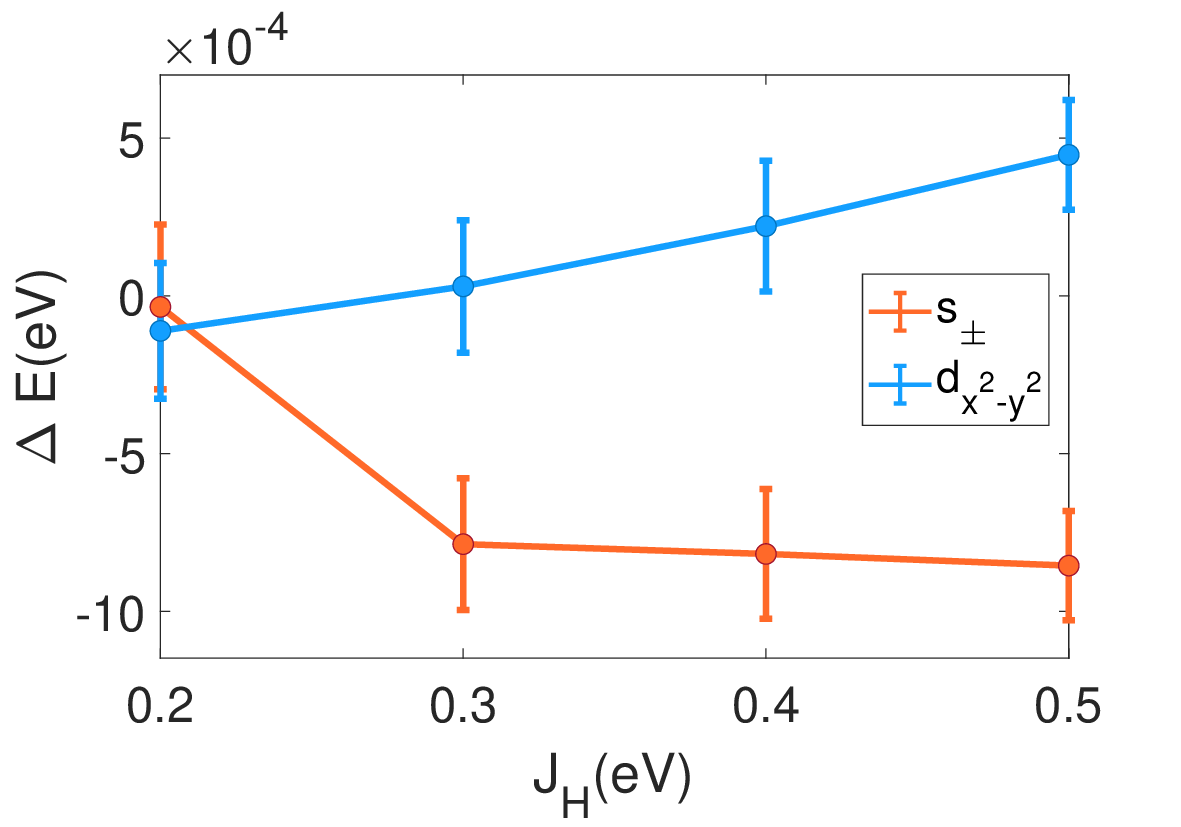}\label{fig:energyjhhubbard}}};
	\node[above left] at (0.35,3.6) {(b)};

\end{tikzpicture}
			\begin{tikzpicture}

		\node[anchor=south west,inner sep=0] (image) at (0,0) {\subfigure{\includegraphics[trim=0 0 20 15,clip,width=0.3\textwidth]{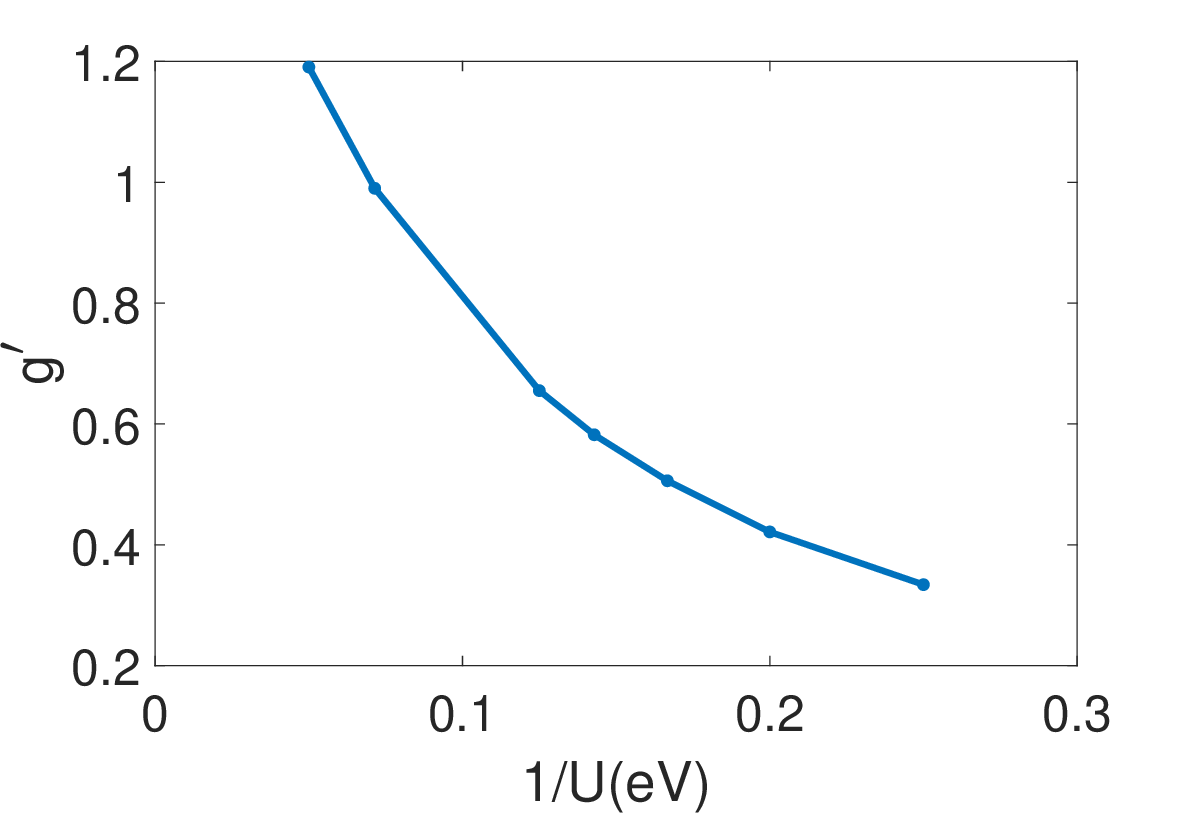}\label{fig:orderg}}};
		\node[above left] at (0.25,3.6) {(c)};
	\end{tikzpicture}
			\begin{tikzpicture}
		\node[anchor=south west,inner sep=0] (image) at (0,0) {\subfigure{\includegraphics[trim=5 5 0 10 ,clip,width=0.3\textwidth]{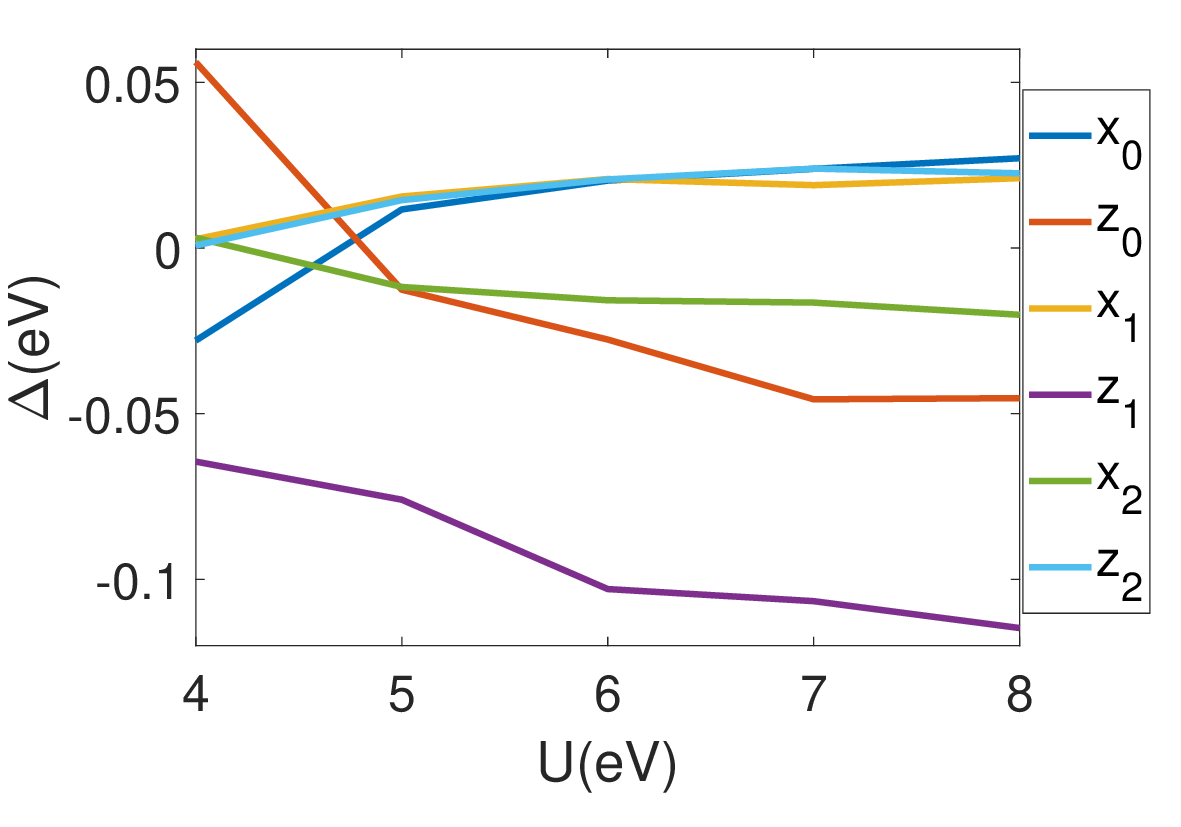}\label{fig:orderUhubbard}}};
		\node[above left] at (0.25,3.6) {(d)};
	\end{tikzpicture}
	\begin{tikzpicture}

		\node[anchor=south west,inner sep=0] (image) at (0,0) {\subfigure{\includegraphics[trim=5 5 10 10,clip,width=0.3\textwidth]{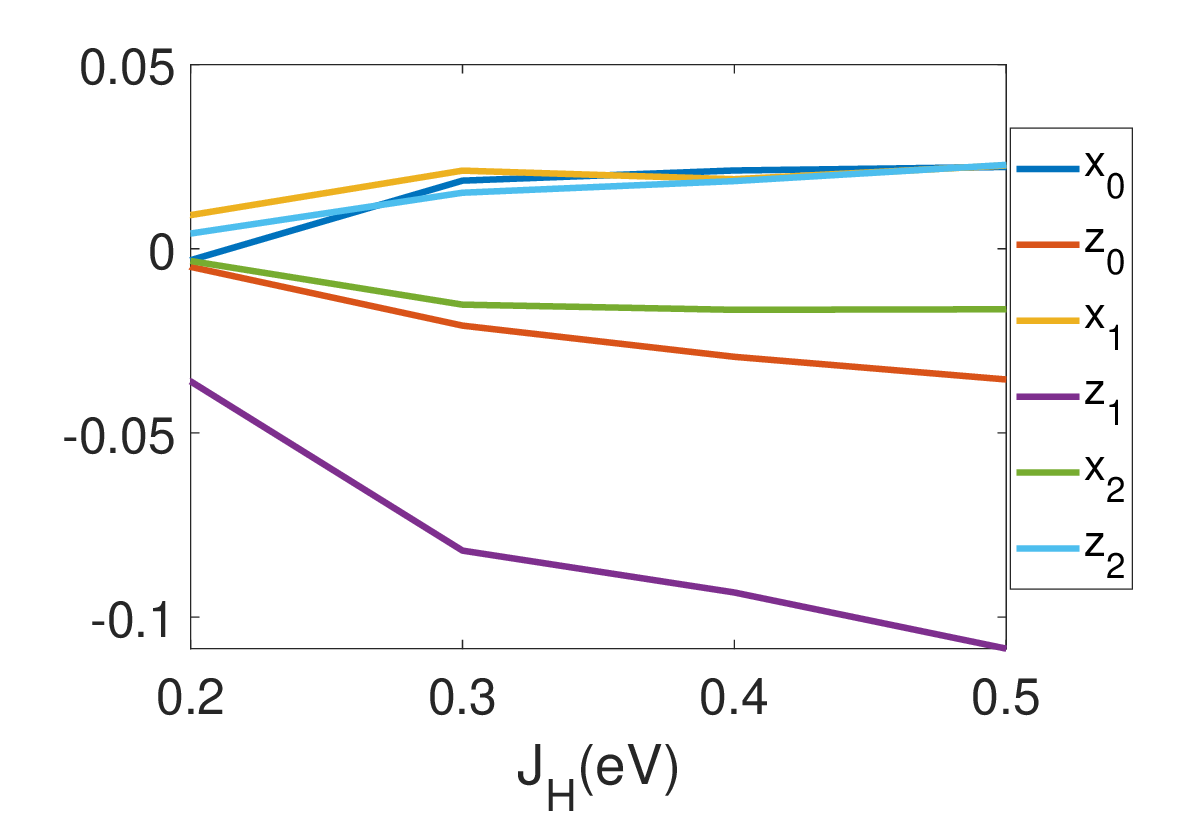}\label{fig:orderjhhubbard}}};
		\node[above left] at (0.25,3.6) {(e)};

	\end{tikzpicture}
	\begin{tikzpicture}
		\node[anchor=south west,inner sep=0] (image) at (0,0) {\subfigure{\includegraphics[trim=0 0 20 15,clip,width=0.3\textwidth]{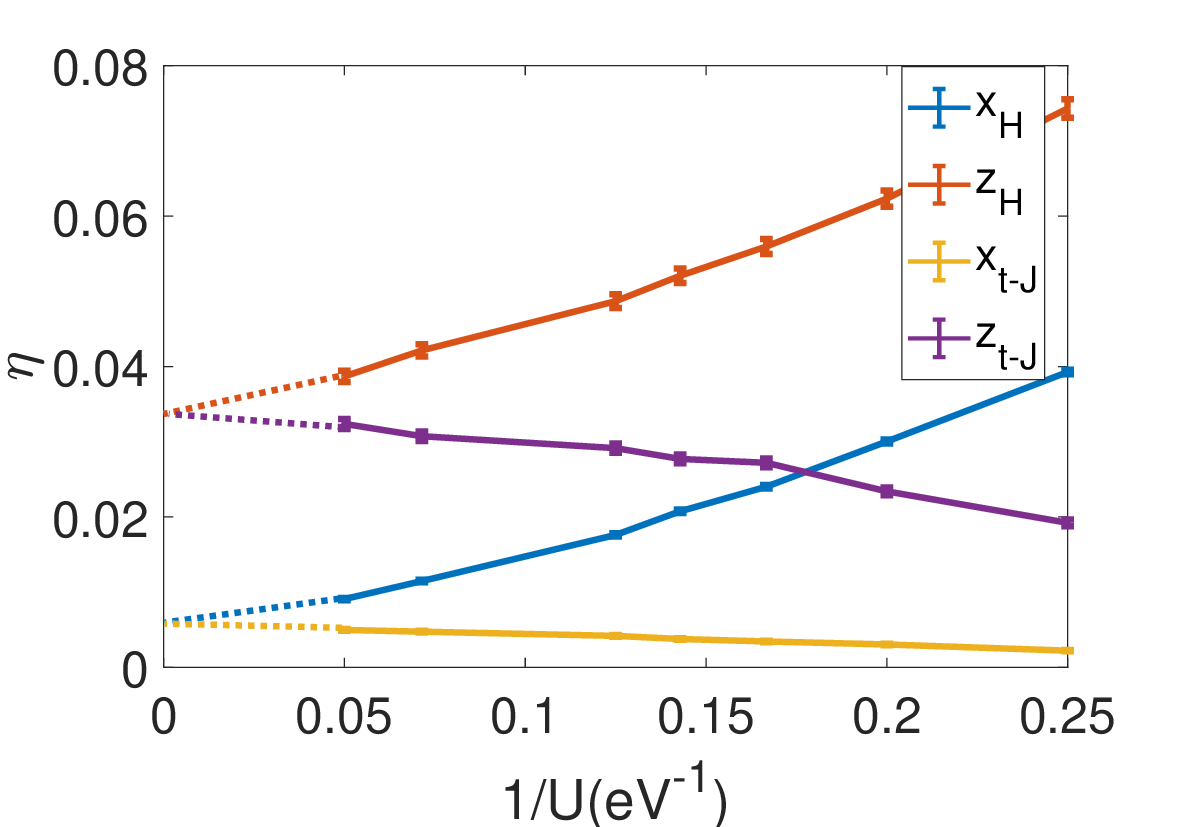}\label{fig:doublehubbard}}};
		\node[above left] at (0.25,3.6) {(f)};

	\end{tikzpicture}
\caption{VQMC results of the Hubbard model. The energies of the $s_\pm$-wave and $d_{x^2-y^2}$-wave pairings relative to the normal state are plotted versus $U$ by fixing $J_H=0.5$~eV in (a), and versus $J_H$ by fixing $U=7$~eV in (b). The Jastrow factor $g'$ enforcing the $1e$ and $2e$ states (see Eq.~\ref{eq:PH}) is plotted versus $1/U$ in (c).
The optimized variational pairing parameters for the $s_\pm$-wave are plotted versus $U$ in (e) and versus $J_H$ in (f), respectively. In (f), the intra-orbital double occupancies $\eta_x$ and $\eta_z$ are plotted versus $1/U$ for both the $t-J$ (denoted by $x_{t-J}$ and $z_{t-J}$) and Hubbard (denoted by $x_{H}$ and $z_H$) models.
\label{fig:hubbard}}
\end{figure*}
\section{VQMC calculations}

In this section, we perform VQMC calculations to explore the superconducting properties of La$_3$Ni$_2$O$_7$.
In VQMC, we take the ansatz that the superconducting ground state $\ket{\Psi_G}$ is given by
\begin{align}
\ket{\Psi_G}=\hat{P}\ket{\Psi_0},
\end{align}
where $\hat{P}$ is a suitable projection operator to be specified below and $\ket{\Psi_0}$ is the many body ground state of a free variational Hamiltonian defined as
\begin{equation}
H_{v}=\sum_{i j a b \sigma}\left[h_{i j}^{a b} c_{i a \sigma}^{\dagger} c_{j b \sigma}+\sigma\left(\Delta_{i j}^{a} c_{i a \sigma}^{\dagger} c_{j a \bar{\sigma}}^{\dagger}+\text { H.c. }\right)\right],
\end{equation}
where $i$ and $j$ denote lattice sites, $a$ and $b$ denote orbitals $x$/$z$, $\sigma=\pm1$ denotes spin.
The hopping integrals $h_{i j}^{a b}$ (including onsite energies) and pairings $\Delta_{ij}^a$ are all variational parameters. In our calculations, we assume $h_{ij}^{ab}=g_{t,ij}^{ab}t_{ij}^{ab}$. For pairing parameters $\Delta_{ij}^{a}$, we consider the $s_{\pm}$-wave and $d_{x^2-y^2}$-wave pairing ansatz, as shown in Fig.~\ref{fig:conswave} and \ref{fig:condwave}, respectively.

We specify the projection operator $\hat{P}$. For the $t-J$ model, we choose
\begin{align}
\hat{P}_{tJ}=\prod_i(Q_{1e,i}+Q_{2e,i})\mathrm{e}^{-\sum_a g_a n_{ia\uparrow}n_{ia\downarrow}}\mathrm{e}^{g_{H}m_{ix}m_{iz}} ,
\end{align}
where $m_{ia}=(n_{ia\uparrow}-n_{ia\downarrow})$.
The first term $(Q_{1e,i}+Q_{2e,i})$ is to project out the $1e$ and $2e$ subspace. In the second term, $g_a$ is the Gutzwiller projection factor to suppress the double occupancy $\eta_a$ for orbital $a$. In the third term, $g_{H}$ is the Jastrow factor to promote the effect of the Hund's coupling.

Different from the $t-J$ model, the Hubbard model needs more careful treatment to obtain superconductivity, since the bare interactions do not favor superconductivity directly (in the mean field sense). Therefore, we need to add more Jastrow factors in the total projection operator giving
\begin{align}
\hat{P}_{H}=&\prod_i\mathrm{e}^{g'(Q_{1e,i}+Q_{2e,i})}\mathrm{e}^{-\sum_a g_a n_{ia\uparrow}n_{ia\downarrow}}\mathrm{e}^{g_{H}m_{ix}m_{iz}} \nonumber\\
&\times\prod_{\langle ij\rangle}\mathrm{e}^{g_1(Q_{0e,i}Q_{2e,j}+Q_{0e,j}Q_{2e,i})}\mathrm{e}^{g_2(Q_{1e,i}Q_{3e,j}+Q_{1e,j}Q_{3e,i})} .  \label{eq:PH}
\end{align}
The first term $g'$ is to promote the $1e$ and $2e$ states since all occupied states are allowed in the Hubbard model. The second and third terms are the same as the $t-J$ model. The last two Jastrow factors $g_1$ and $g_2$ are added to promote the doublon-holon processes involving $0e$ and $3e$ states, respectively, which are responsible for the super-exchange processes. In the above defined $\hat{P}_{tJ}$ and $\hat{P}_{H}$, all the $g$ factors are positive.

By VQMC, all the variational parameters $h_{ij}^{ab}$, $\Delta_{ij}^{a}$, and the $g$-factors are optimized to minimize the total energy $\langle \psi_G|H_{H,tJ}|\psi_G\rangle$ for the Hubbard and $t-J$ models, respectively, using the stochastic reconfiguration technique. More technical details about the VQMC can be found in the appendix \ref{VQMC_details}.
In practice, the periodic-antiperiodic boundary condition is adopt to avoid the open-shell problem. Most of our VQMC simulations are performed on the $L\times L\times2$ lattice with $L=6$, but larger sizes up to $L=12$ have been checked, as shown in the appendix \ref{results}, to ensure the finite size effect is acceptably small.

\subsection{VQMC results of the $t-J$ model}
We first present the results of the $t-J$ model. The optimized ground state energies of the $s_{\pm}$-wave and $d_{x^2-y^2}$-wave pairings relative to the normal state are plotted versus $U$ (by fixing $J_H=0.5$~eV) in Fig.~\ref{fig:energyU} and versus $J_H$ (by fixing $U=7$~eV) in Fig.~\ref{fig:energyjh}, respectively. In the simulated parameter regime, the $s_{\pm}$-wave pairing always saves more energy than the $d_{x^2-y^2}$-wave pairing, which is consistent with the early RMFT result \cite{PhysRevB.108.L140505} indicating the robustness of the $s_\pm$-wave pairing in La$_3$Ni$_2$O$_7$.
With increasing $U$, the energy gain is found to be reduced, indicating the important role of the super-exchange $J=4(t_{ij}^{aa})^2/(U-J_H)$ in the superconducting mechanism in the large $U$ limit, which is somewhat similar to the single-orbital $t-J$ model. On the other hand, the energy gain is also found to grow up with increasing $J_H$. These results indicate that both the super-exchange and the Hund's coupling promote the superconductivity in La$_3$Ni$_2$O$_7$ at least in the strong coupling (large $U$) limit.

We then examine the optimized pairing parameters for the $s_\pm$-wave pairing as shown in Fig.~\ref{fig:orderU} versus $U$ and in Fig.~\ref{fig:orderjh} versus $J_H$. Other variational parameters can be found in appendix \ref{results}. Here, we focus on the pairing parameters.
Onsite pairing parameters $\Delta_{ii}^x$ (denoted by $x_0$) and $\Delta_{ii}^z$ (denoted by $z_0$) are anti-phase due to the pair-hopping interaction $J_P$. Note that the large $x_0/z_0$ does not necessarily lead to large order parameter $\langle c_{i,x/z,\downarrow} c_{i,x/z,\uparrow}\rangle$ because of the projection which suppresses the double occupancy (see below). Besides the onsite pairings, the dominant pairing is on the vertical bond $\Delta_{ij}^z$ (denoted by $z_1$), which is much larger than $\Delta_{ij}^x$ ($x_1$), indicating the leading role of 3d$_{3z^2-r^2}$ orbital in the $s_\pm$-wave pairing. As a comparison, the planar pairings $z_2$ and $x_2$ are relatively small. These results are also qualitatively consistent with the early RMFT calculations \cite{PhysRevB.108.L140505}.

\subsection{VQMC results of the Hubbard model}
We turn to the more generic Hubbard model. As explained above, we need to add more Jastrow factors including $g'$ (to enforce $1e$ and $2e$ states), $g_1$ and $g_2$ (to enforce doublon-holon process on the vertical and planar bonds). After VQMC optimization, our results are shown in Fig.~\ref{fig:hubbard}. The energy gains of the $s_\pm$ and $d_{x^2-y^2}$ pairings relative to the normal state are plotted versus $U$ (fixing $J_H=0.5$~eV) in Fig.~\ref{fig:energyUhubbard} and versus $J_H$ (fixing $U=7$~eV) in Fig.~\ref{fig:energyjhhubbard}. We find the $s_\pm$-wave pairing always saves more energy than the $d_{x^2-y^2}$-wave pairing as long as $J_H>0.2$~eV (very likely in La$_3$Ni$_2$O$_7$).
This is one of the main results of the present work. It connects the weak-$U$ limit (as studied by FRG) and strong-$U$ limit ($t-J$ model as studied by RMFT), indicating the $s_\pm$-wave pairing is the most likely candidate of the pairing symmetry in La$_3$Ni$_2$O$_7$.
Moreover, more energy gain is obtained for the $s_\pm$-wave pairing with decreasing $U$ and/or increasing $J_H$, similar to the above $t-J$ model. This is consistent with the physical picture that both the super-exchange and Hund's coupling can promote the antiferromagnetic correlation on the vertical bond, which further induces the $s_\pm$-wave superconductivity.

Then we analyze the optimized variational parameters. In Fig.~\ref{fig:orderg}, the Jastrow factor $g'$ (enforcing the $1e$ and $2e$ states) is found to grow up quickly as increasing $U$, confirming the validity of the $t-J$ model (living in the $1e$ and $2e$ subspace). In Fig.~\ref{fig:orderUhubbard} and \ref{fig:orderjhhubbard}, we plot the optimized variational pairing parameters for the $s_\pm$-wave versus $U$ and $J_H$, respectively. Other variational parameters and Jastrow factors can be found in appendix \ref{results}. The leading pairing component is $z_1$ (between 3d$_{3z^2-r^2}$ orbitals and on the vertical bond), which is consistent with the $t-J$ model and confirms the leading role of the 3d$_{3z^2-r^2}$ orbital in the superconducting pairing. The subleading pairing components are $z_0$ and $x_0$ (onsite pairings) opposite signs, which is consistent with the atomic repulsive pair-hopping interaction.

Finally, we present the intra-orbital double occupancies $\eta_{x}=\langle Q_{2e,ix}\rangle$ and $\eta_z=\langle Q_{2e,iz}\rangle$ obtained in both the Hubbard and $t-J$ models, as shown in Fig.~\ref{fig:doublehubbard}. For the Hubbard model, both $\eta_x$ and $\eta_z$ drops with increasing $U$, reflecting the effect of the Hubbard $U$ on suppressing the double occupancy. Instead, for the $t-J$ model, both $\eta_x$ and $\eta_z$ grow up with increasing $U$ rather than drops down. This seemingly paradox can be reconciled by the fact that the $t-J$ model already lives in the vicinity of $U=U'=\infty$ such that intra-orbital and inter-orbital double occupancies have no difference. We find in the limit of $U\to\infty$, both the Hubbard and $t-J$ models give the same double occupancies, which can be taken as a benchmark of the correctness of our VQMC simulations on both two models. Note that $\eta_z$ keeps a finite value even as $U\to\infty$. This is quite different from the half-filled Hubbard model, for which the local moment can be well defined by the vanishing of double occupancy. But for La$_3$Ni$_2$O$_7$, due to the non-integer filling fraction $1.5$ per site, no well defined local moment exists even in the strong-coupling limit. This sheds light on the itinerant versus local moment picture of the electrons in this material.

\section{Summary}

In this work, we investigate the pairing symmetry in La$_3$Ni$_2$O$_7$ under pressure by applying VQMC on bilayer two-orbital $t-J$ and Hubbard models. We find the $s_\pm$-wave pairing is robust in most parameter regimes, which is consistent with the early FRG from weak to moderate correlation limit and RMFT for strong-coupling limit. The leading pairing component is found to the one between 3d$_{3z^2-r^2}$ orbitals on the vertical bond, indicating the important role of the 3d$_{3z^2-r^2}$ orbital in superconducting La$_3$Ni$_2$O$_7$. We also find the intra-orbital double occupancies for the two orbitals remain finite even in the strong-coupling limit, shedding light on the itinerant versus local moment picture of the electrons in this material.

	\begin{acknowledgments}
		This work is supported by National Key R\&D Program of China (Grant No. 2022YFA1403201 and No. 2024YFA1408100) and National Natural Science Foundation of China (Grant No. 12374147, No. 12274205 and No. 92365203).
	\end{acknowledgments}

\appendix

\section{\label{VQMC_details} Details of VQMC method}
	We use $\mathbf{X}$ to represent the set of variational parameters $(x_1,x_2\cdots x_\mu)$, and the variational free Hamiltonian can be written in a matrix form (in Nambu space) as
	\begin{equation}
		H_{v} = \mathbf{\phi^\dagger h(X) \phi},
	\end{equation}
	where $\mathbf{\phi}^\dagger = (c_{1\uparrow}^\dagger, c_{2\uparrow}^\dagger \cdots c_{N\uparrow}^\dagger,c_{1\downarrow},c_{2\downarrow} \cdots c_{N\downarrow})$, $N=4N^{\prime}$ (two orbitals and two spins) with $N^{\prime}$ the number of unitcells, and $\mathbf{h(X)}$ is the Hamiltonian matrix in the basis of $\mathbf{\phi}$. In the following,  we will not write out $\mathbf{X}$ explicitly for simplicity. By diagonalization,
	\begin{equation}
		H_{v} = \mathbf{\phi}^\dagger \mathbf{U U^\dagger h(X) U U^\dagger} \mathbf{\phi} = \mathbf{\psi}^\dagger \mathbf{E} \mathbf{\psi} = \sum_\alpha \lambda_\alpha \psi_\alpha^\dagger \psi_\alpha,
	\end{equation}
	where the diagonal matrix $\mathbf{E} = \text{diag}(\lambda_1 \cdots \lambda_{2N})$ contains $2N$ eigenvalues of $\mathbf{h}$, and $\mathbf{U}$ is the unitary transformation matrix composed by eigenvectors $(\mu_{\alpha},\nu_{\alpha})^t$ corresponding to $\lambda_\alpha$. We suppose $\lambda_n \leq 0$ with $n\leq N$, and construct two $N \times N$ matrices
	\begin{equation}
		V_1=(\mu_1, \mu_2 \cdots \mu_{N}), \quad V_2=(\nu_1, \nu_2 \cdots \nu_{N}),
	\end{equation}
	from which we can write down the ground state of $H_{v}$ as
	\begin{equation}
		\ket{\Psi_0} = \left(\sum_{ij}c_{i\uparrow}^{\dagger}A_{ij}c_{j\downarrow}^{\dagger}\right)^{N_e/2}\ket{0} .
	\end{equation}
    Here, $\ket{0}$ is the vacuum state, and $N_e=3N/4$ is the total number of electrons, and
	\begin{equation}
		A=V_1V_2^{-1}.
	\end{equation}
The ansatz of the variational ground state is
\begin{align}
\ket{\Psi_G}=\hat{P}\ket{\Psi_0} ,
\end{align}
where $\hat{P}$ is a projection operator containing suitable Jastrow factors as specified in the main text. In the basis of real space many-body configuration $\{\ket{R}\}$, the variational ground state can be expanded as
\begin{equation}
	\ket{\Psi_{G}} = \sum_R \ket{R}\braket{R|\Psi_{G}} = \sum_R f_R\text{det}(A_R) \ket{R},
\end{equation}
where $f_R$ is the projection factor for a given configuration $\ket{R}$, and $\text{det}(A_R)$ is the determinant of the matrix $A_R$.
The total energy of the Hubbard or $t-J$ model can be calculated as
\begin{align}
E&=\frac{\braket{\Psi_G|H|\Psi_G}}{\braket{\Psi_G|\Psi_G}} \nonumber\\
&=\frac{ \sum_{RR'} f_Rf_{R'} \text{det}(A_R^*) \text{det}(A_{R'}) \braket{R|H|R'} }{ \sum_R f_R^2 |\text{det}(A_R)|^2 } \nonumber\\
&=\frac{ \sum_{R} f_R^2|\text{det}(A_R)|^2 \sum_{R'}\frac{f_{R'}}{f_R} \frac{ \text{det}(A_{R'})}{\text{det}(A_R)} \braket{R|H|R'} }{ \sum_R f_R^2 |\text{det}(A_R)|^2 } \nonumber\\
&=\left\langle \sum_{R'}\frac{f_{R'}}{f_R} \frac{ \text{det}(A_{R'})}{\text{det}(A_R)} \braket{R|H|R'} \right\rangle,
\end{align}
where we have defined the Monte Carlo average $\langle \cdots\rangle $
with the configuration weight $w_R=f_R^2|\text{det}(A_R)|^2$, i.e.
\begin{align}
\braket{\cdots}=\frac{\sum_R (\cdots) w_R}{\sum_R w_R} .
\end{align}
Next, to optimize the variational parameters $\mathbf{X}$, we adopt the stochastic reconfiguration method \cite{PhysRevB.71.241103,Cyrus2007} to update $x_\mu$.
In the usual gradient descent method,
\begin{align}
dx_\mu = -\frac{\partial E}{\partial x_\mu} dt ,
\end{align}
where $dt$ is an artificial time step. After updating all variational parameters, we have
\begin{equation}
	\ket{\Psi_{G}} \rightarrow  \ket{\Psi_{G}} + \sum_\mu dx_\mu\ket{\mu},
\end{equation}
	where $\ket{\mu} \equiv \partial\ket{\Psi_{G}}/\partial x_\mu$. However, this updating strategy is not always stable since the states {$\ket{\mu}$} are not be orthogonal to each other in general such that the variational parameters are not independent and the update in $x_\mu$ could have a huge influence on $x_\nu$. As a remedy, we can choose a set of orthogonal basis $\{\ket{a}\}$ such that
\begin{equation}
dx_a=-\frac{\partial E}{\partial x_a} dt .
\end{equation}
and
	\begin{equation}
		\sum_a dx_a\ket{a} = \sum_\mu dx_\mu\ket{\mu}.
	\end{equation}
Left-multiplying with $\bra{\nu}$ on the two sides, we have
\begin{align}
\sum_a dx_a \braket{\nu|a}=\sum_\mu \braket{\nu|\mu}  dx_\mu \equiv\sum_\mu g_{\nu\mu}dx_\mu,
\end{align}
where we have defined $g_{\nu\mu}\equiv\braket{\nu|\mu}$.
Then, left-multiplying with $(g^{-1})_{\mu'\nu}$, summing over $a,\mu,\nu$, and then replacing $\mu'$ with $\mu$, we obtain the revised update equation
\begin{align}
dx_{\mu}= -\sum_\nu (g^{-1})_{\mu\nu} \frac{\partial E}{\partial x_\nu} dt.
\end{align}
In general, the matrix $g$ could have eigenvalues approaching zero, which  could cause instability when calculating the inverse of $g$. We can add a small positive number $\varepsilon$ to the diagonal elements of $g$, i.e., $g_{\mu\mu} \rightarrow (1 + \varepsilon)g_{\mu\mu}$.
Now we can update $x_\mu$ as long as we obtain the matrix $g_{\mu\nu}$ and $\partial_\nu E$, which can be calculated in the basis of $\ket{R}$ as
\begin{align}
	g_{\mu\nu} &= \braket{l^*_\mu l_\nu} - Re\braket{l_\mu}Re\braket{l_\nu} \nonumber\\ &	\quad + i[Im\braket{l_\mu}Re\braket{l_v} - (\mu \leftrightarrow \nu)]\label{g}, \\
    \partial_\nu E &= \braket{hl_\nu} - E\braket{l_\nu} + c.c.,\label{e}
\end{align}
where
\begin{align}
l_\mu(R) &\equiv \partial_\mu \rm{ln}\left[f_R\text{det}(A_R)\right] \nonumber \\
&= \rm{Tr}(A^{-1}_R \partial_\mu A_R)+ \partial_\mu \ln(f_R) .
\end{align}

	\begin{figure}[b]
	\centering
	\hspace{-25pt}
	\begin{tikzpicture}

		\node[anchor=south west,inner sep=0] (image) at (0,0) {\subfigure{\includegraphics[trim=5 18 35 10,clip,width=0.3\textwidth]{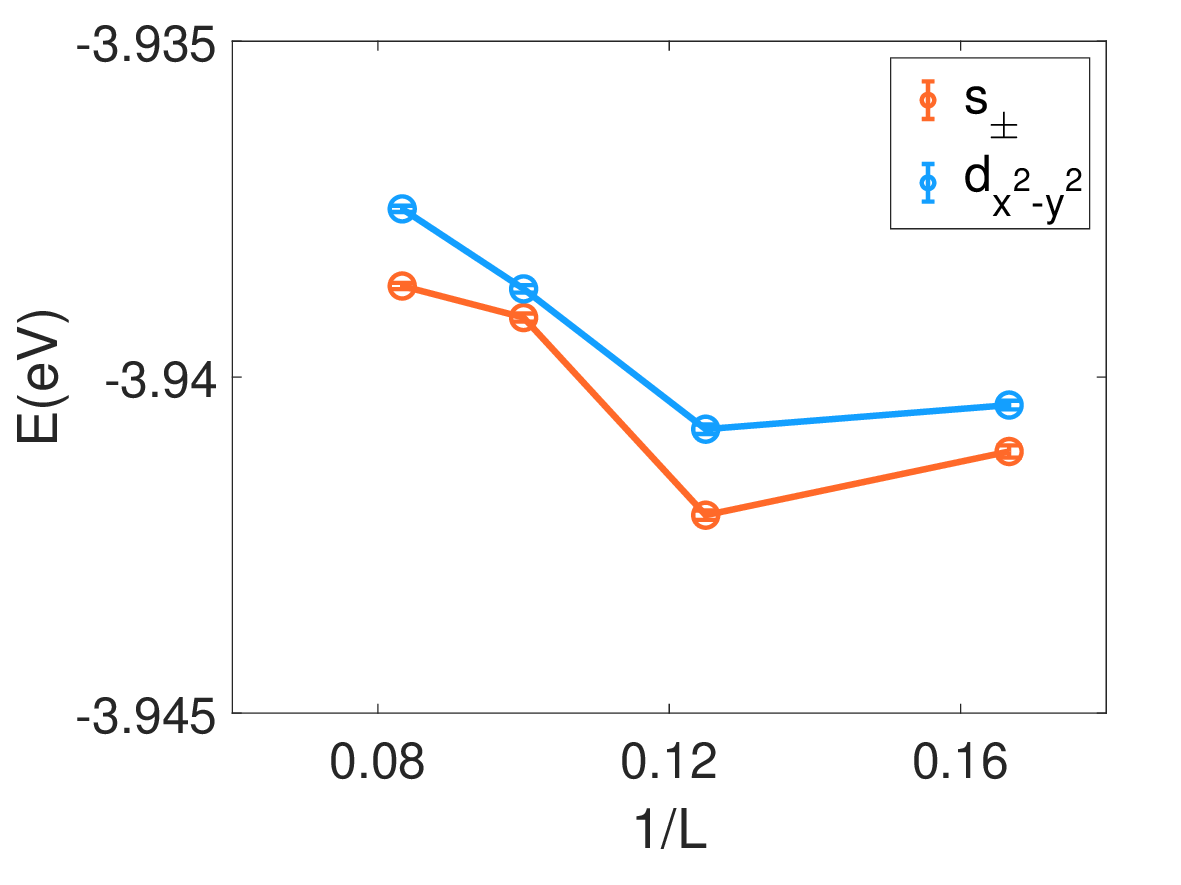}\label{fig:fse}}};

	\end{tikzpicture}

	\caption{ Finite size effect with lattice size $L=6$ to $12$ for the energies of the $s_\pm$-wave and $d_{x^2-y^2}$-wave pairings at $U=7$~eV and $J_H=0.5$~eV.}
	\label{fig:9}
\end{figure}

\begin{figure}
		\centering
		\hspace{5pt}
		\begin{tikzpicture}

			\node[anchor=south west,inner sep=0] (image) at (0,0) {\subfigure{\includegraphics[trim=80 0 180 10,clip,width=\linewidth]{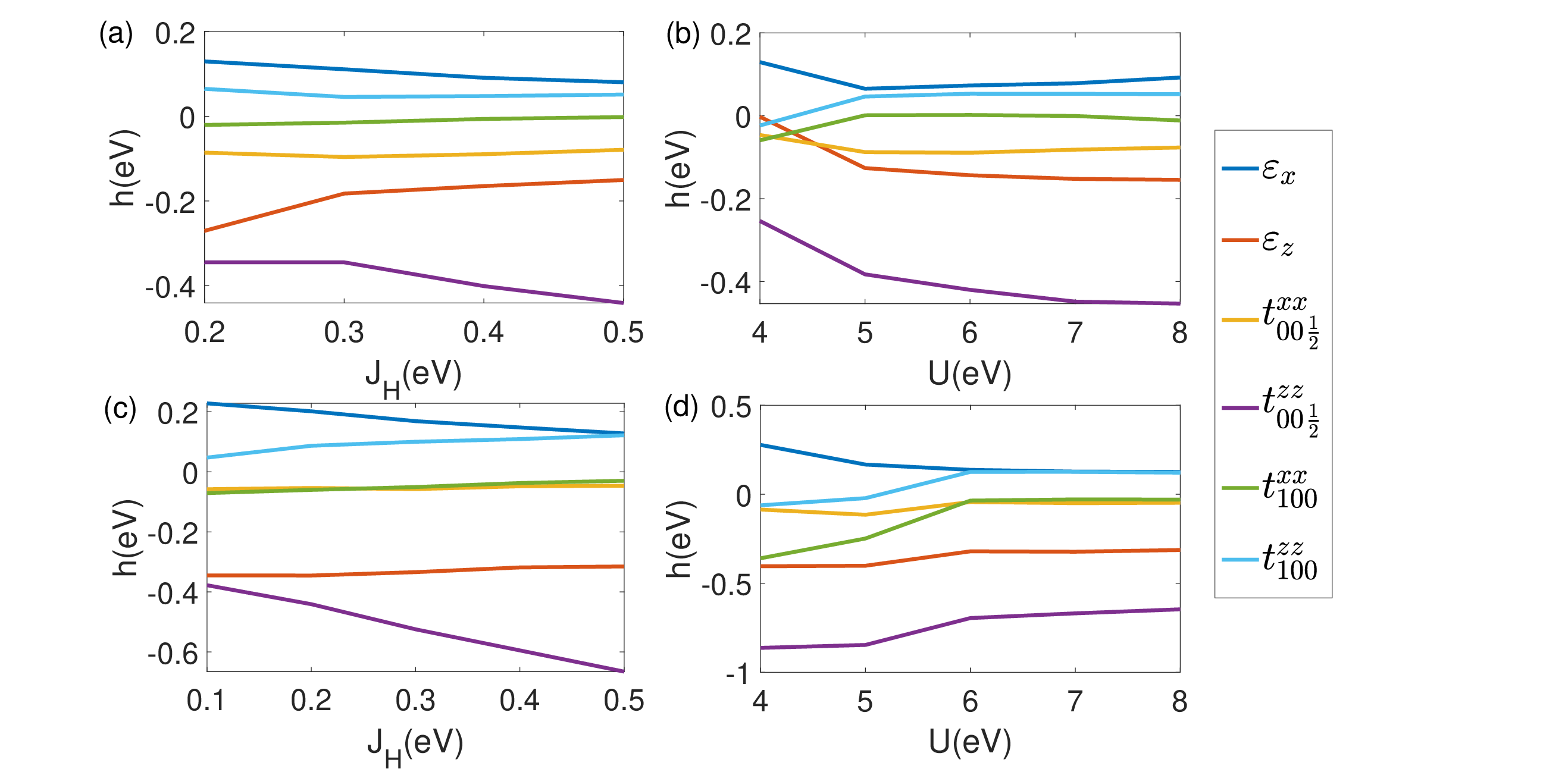}\label{fig:hop}}};

		\end{tikzpicture}

		\caption{ The optimized variational hopping and onsite energy parameters for the Hubbard model are plotted versus $J_H$ at $U=7$~eV in (a), and versus $U$ at $J_H=0.5$~eV in (b). (c,d) are similar to (a,b) but for the $t-J$ model.}
		\label{fig:7}
	\end{figure}

	\begin{figure}
		\centering
		\hspace{10pt}
		\begin{tikzpicture}

			\node[anchor=south west,inner sep=0] (image) at (0,0) {\subfigure{\includegraphics[trim=100 0 150 0,clip,width=\linewidth]{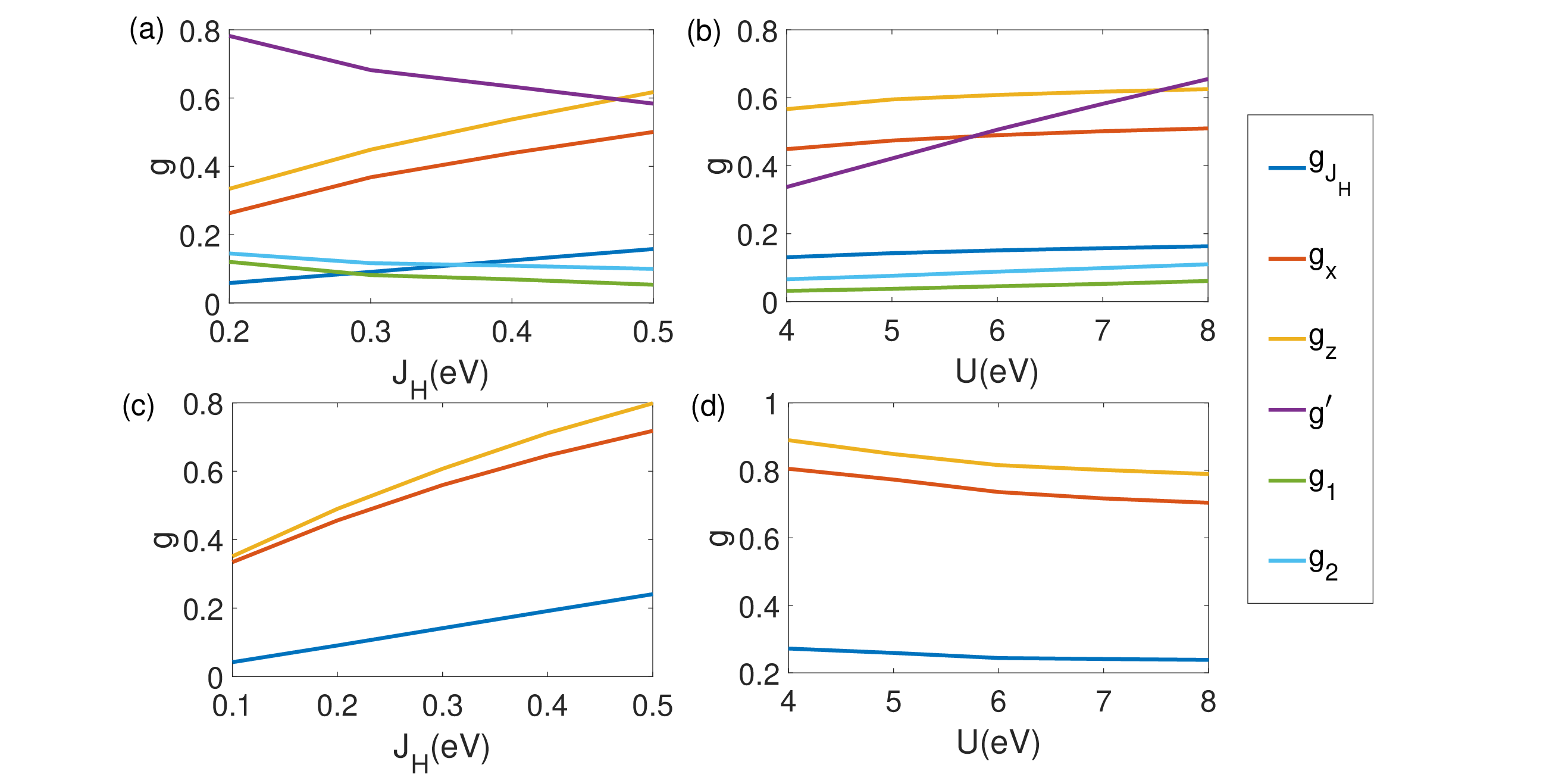}\label{fig:jas}}};

		\end{tikzpicture}

		\caption{The optimized Jastrow factors for the Hubbard model are plotted versus $J_H$ at $U=7$~eV in (a), and versus $U$ at $J_H=0.5$~eV in (b). (c,d) are similar to (a,b) but for the $t-J$ model. }
		\label{fig:8}
	\end{figure}

\section{\label{results} More VQMC results}

In this section, we supplement more results of our VQMC simulations.
In Fig.~\ref{fig:9}, we plot the energies of the $s_\pm$-wave and $d_{x^2-y^2}$-wave pairings for the $t-J$ model on different lattice size $L$ (from $6$ to $12$) to examine the finite size effect. We find the energies does not change with $L$ significantly, in particular the relative energies between the two pairing candidates. This indicates that $L=6$ is large enough and reliable to identify the pairing symmetry.
We next present more optimized variational parameters.
In Fig.~\ref{fig:7}, we present the variational hoppings and onsite energies in the free variational Hamiltonian $H_v$ relative to the bare values in the tight-binding Hamiltonian $H_0$.
The inter-layer hopping between the 3d$_{3z^2-r^2}$ orbitals $h_{00\frac12}^{zz}$ keeps to have the largest amplitude, indicating its important role (e.g. for super-exchange) even in the strong-coupling limit.
In Fig.~\ref{fig:8}, we present the optimized Jastrow factors. In particular, $g_{x,z}$ (to suppress intra-orbital double occupancy) grows up with increasing $U$, consistent with the double occupancies $\eta_{x,z}$ as shown in the main text.

\bibliography{ref}

\end{document}